\documentclass[12pt]{article}
\pdfoutput=1
\usepackage[utf8]{inputenc}
\usepackage{amsmath,amssymb,graphicx,epsfig,color,float,cancel,cite}
\usepackage[titletoc,title]{appendix} 
\usepackage[export]{adjustbox}
\usepackage{multicol,tabu,multirow}
\usepackage{arydshln}
\usepackage{hyperref}
\usepackage{slashed}                                                                                                                                                                                                                                                                                                                                                                                                                                                                                                                                                                                                                                                                                                                                                                                                                                                                                                                                                                                                                                                                                                
\numberwithin{equation}{section}

\def\nn{\nonumber}

\def\nn{\nonumber}
\def\bea{\begin{eqnarray}}
\def\eea{\end{eqnarray}}
\def\bal#1\eal{\begin{align}#1\end{align}}
\newcommand{\beq}{\begin{equation}}
\newcommand{\eeq}{\end{equation}}
\newcommand{\bseq}{\begin{subequations}}
\newcommand{\eseq}{\end{subequations}}
\allowdisplaybreaks
\begin{document}
\thispagestyle{empty}
\begin{center}
\vspace*{1cm}
{\LARGE\bf Explaining the $R_K$  and $R_{K^*}$ anomalies \\}
\bigskip
{\large Diptimoy Ghosh}
\\
\bigskip 
\bigskip
{Department of Particle Physics and Astrophysics, Weizmann Institute of Science, 
\\ Rehovot 76100, Israel. 
}
\end{center}
\bigskip 
\vspace*{1cm}
\begin{center} 
{\Large\bf Abstract} 
\end{center}
\vspace*{-0.35in}
\begin{quotation}
\noindent 
Recent LHCb results on $R_{K^*}$, the ratio of the branching fractions of $B \to K^* \mu^+ \mu^-$ to that of $B \to K^* e^+ e^-$, 
for the dilepton invariant mass bins $q^2 \equiv m_{\ell\ell}^2 = [0.045 - 1.1]$ GeV$^2$ and  $[1.1 - 6]$  GeV$^2$ show 
approximately $2.5 \sigma$ deviations from the corresponding Standard Model prediction in each of the bins. This, when 
combined with the measurement of $R_K \, (q^2=[1-6]\, \rm GeV^2)$, a similar ratio for the decay to a pseudo scalar meson, 
highly suggests for lepton non-universal new physics in semi-leptonic $B$ meson decays. In this work, we perform a model 
independent analysis of these potential new physics signals and identify the operators that do the best job in satisfying all 
these measurements. We show that heavy new physics,  giving rise to $q^2$ independent local 4-Fermi operators of scalar, 
pseudo-scalar, vector or axial-vector type, is unable to explain all the three measurements simultaneously, in particular $R_{K^*}$ in the 
bin [0.045 - 1.1], within their experimental $1\sigma$ regions. We point out the possibility to explain $R_{K^*}$ in the 
low bin by an additional light ($\lesssim 20 \, \rm MeV$) vector boson with appropriate coupling strengths to ($\bar b \, s$) and 
($\bar e \, e$).
\end{quotation}
\bigskip
%
%
\vfill
%
%
\bigskip
\hrule
\hspace*{4.8cm} diptimoy.ghosh@weizmann.ac.il

\newpage 
\section{Introduction}

The LHCb collaboration has recently reported hints of new physics (NP) in lepton flavour non-universal observables $R_K$ and 
$R_{K^*}$, 
\bea
R_{K^{(*)}} =\frac{\mathcal{B} \left( B \to K^{(*)} \mu^+ \mu^-\right)}{ \mathcal{B} \left( B \to K^{(*)} e^+ e^-\right)} \, .
\eea
While the result for $R_K$ was presented only in the dilepton invariant mass squared, $q^2 \in$ [1 - 6] GeV$^2$, $R_{K^*}$ has been measured 
in two bins, [0.045 - 1.1] GeV$^2$ and  [1.1 - 6]  GeV$^2$. The experimental results are summarised in table~\ref{exp-data}. 

\begin{table}[!h]
\begin{center}
\begin{tabular}{|c|cr|cr|} 
\hline 
Observable & SM prediction &  & Measurement  &  \\
\hline 
$R_K : q^2 = [1,6] \, \text{GeV}^2$  & $1.00 \pm 0.01 $&  \cite{Descotes-Genon:2015uva,Bordone:2016gaq} &  $0.745^{+0.090}_{-0.074} \pm 0.036$  & \cite{Aaij:2014ora} \\
\hline
$R_{K^*} ^{\rm low}: q^2 = [0.045,1.1] \, \text{GeV}^2$  & $0.92 \pm 0.02$  &  \cite{Capdevila:2017bsm} &  $0.660^{+0.110}_{-0.070} \pm 0.024$  & 
\cite{LHCb} \\
\hline
$R_{K^*}^{\rm central} : q^2 = [1.1,6] \, \text{GeV}^2$  & $1.00 \pm 0.01 $&  \cite{Descotes-Genon:2015uva,Bordone:2016gaq} &  $0.685^{+0.113}_{-0.069} \pm 0.047$  & 
\cite{LHCb} \\
\hline
\hline
$\mathcal{B} \left(B_s \to \mu^+ \mu^-\right)$ 
& $ \left( 3.57 \pm 0.16 \right) \times 10^{-9}$ & \cite{Bobeth:2013uxa,Fleischer:2017ltw}
& $\left( 3.00 \pm 0.5   \right) \times 10^{-9}$ & \cite{Chatrchyan:2013bka,Aaij:2017vad,Fleischer:2017ltw} \\
\hline
$\mathcal{B} \left( B_s \to e^+ e^-\right)$ 
& $ \left( 8.35 \pm 0.39 \right) \times 10^{-14}$ & \cite{Bobeth:2013uxa,Fleischer:2017ltw}
& $ < 2.8  \times 10^{-7}$ & \cite{Aaltonen:2009vr} \\
\hline
\end{tabular}
\caption{The observables used in our analysis along with their SM predictions and experimental measurements. Note that the QED 
corrections to $R_K$ and $R_{K^*}$ in the bin $q^2 = [1,6] \, \text{GeV}^2$ were first calculated in \cite{Bordone:2016gaq}. However, 
no such calculation exists for $R_{K^*}$ in the bin $q^2 = [0.045,1.1] \, \text{GeV}^2$. \label{exp-data}}
\end{center}
\end{table} 
While the deviations from the Standard Model (SM) in the individual ratios are only at the level of $2.2\sigma - 2.5\sigma$, the combined 
deviation (the exact number depends on how one combines the 3 results) is large enough to look for NP explanations
\footnote{Similar anomalies have also been observed in the charged current decays $(B \to D^{(*)} \tau \nu /B \to D^{(*)} \ell \nu)$ that call for lepton non-universal 
new physics. See \cite{Bardhan:2016uhr,Alonso:2016oyd} for some recent studies.}. For recent studies, see 
\cite{Capdevila:2017bsm,Altmannshofer:2017yso,DAmico:2017mtc,Hiller:2017bzc,Geng:2017svp,Ciuchini:2017mik,Celis:2017doq,Becirevic:2017jtw}.

At the quark level, the decays $B \to K^{(*)} \mu^+ \mu^-$ proceed via $b \to s$ flavour changing neutral current (FCNC) transitions. 
These decays are particularly interesting because they are highly suppressed in the SM and many extensions of the SM are capable of 
producing measurable effects beyond the SM. In particular, the three body decay $B \to K^{*} \mu^+ \mu^-$ offers a large number
of observables in the angular distributions of the final state particles, hence providing a lot of opportunities to test the SM, see for example, 
\cite{Hiller:2003js,
Altmannshofer:2008dz,Alok:2009tz,Alok:2010zd,Alok:2011gv,DescotesGenon:2011yn,Altmannshofer:2011gn,Matias:2012xw
,DescotesGenon:2012zf,Lyon:2013gba,Descotes-Genon:2013wba,Altmannshofer:2013foa,Buras:2013qja,Datta:2013kja,
Ghosh:2014awa,Queiroz:2014pra,Mandal:2014kma,Greljo:2015mma,
Gripaios:2015gra,Barbieri:2015yvd,Sahoo:2015wya,Sahoo:2016pet,Feruglio:2016gvd,Barbieri:2016las,GarciaGarcia:2016nvr,Megias:2016bde,Bhattacharya:2016mcc,Bhatia:2017tgo,Megias:2017ove,Datta:2017pfz} 
and references therein for related studies. 

The individual branching ratios $\mathcal{B} \left( B \to K^{(*)} \mu^+ \mu^- \right)$ and $\mathcal{B} \left( 
B \to K^{(*)} e^+ e^- \right)$ are predicted with comparatively larger hadronic uncertainties in the SM. However, 
their ratio is a theoretically clean observable and predicted to be close to unity in the SM. This is in contrast to some of the 
angular observables (for example, $P_5^\prime$) where considerable debate exists surrounding the issue 
of theoretical uncertainty due to (unknown) power corrections to the factorization framework and non-local charm loops, see for example,  
\cite{Ball:2006eu,Khodjamirian:2010vf,Dimou:2012un,Jager:2012uw,Lyon:2013gba,Lyon:2014hpa,Ciuchini:2015qxb,
Jager:2014rwa,Capdevila:2017ert}. 
Hence the observed deviation from the SM might be (at least partly) resolved once these corrections are 
better understood.

Therefore, in this work we will only consider the theoretically clean observables $R_{K^{(*)}}$ listed in table~\ref{exp-data}. 
Additionally, we also consider the branching ratios of the fully leptonic decays $B_s \to \mu^+ \mu^-$ and $B_s \to e^+ e^-$, 
as they are very well predicted in the SM. 

The paper is organised as follows. In the next section, we show the complete set of operators at the dimension 6 level for 
$b \to s \, \ell \, \ell$ transition. In section \ref{results} we discuss in detail how these various operators perform in explaining the  
$R_{K^{(*)}}$ anomalies, and point out the possibility of explaining $R_{K^{*}}$ in the low $q^2$ bin by a very light gauge boson. 
We close in section \ref{conclusion} with a brief summary.  

\section{$b \to s$ effective Hamiltonian}

The effective Hamiltonian for $b \rightarrow s$ transitions in the Standard Model is given by
\begin{equation} \label{eq:heff}
{\cal H}_{\mathrm{eff}} = -\frac{4G_F}{\sqrt{2}} \left(\lambda_t^{(s)} {\cal{H}}_{\mathrm{eff}}^{(t)} + 
\lambda_u^{\mathrm{(s)}} {\cal{H}}_{\mathrm{eff}}^{(u)}\right) + \rm h.c. \, ,
\end{equation}
with the CKM matrix combinations $\lambda_q^{(s)}=V_{qb} V_{qs}^{*}$, and
\begin{eqnarray}
{\cal{H}}_{\mathrm{eff}}^{(t)} &=& C_1 {\cal{O}}_1^c + C_2 {\cal{O}}_2^c + \sum_{i=3}^{6} C_i {\cal{O}}_i + 
\sum_{i=7}^{10} C_i {\cal{O}}_i \nonumber \, , \\
 {\cal{H}}_{\mathrm{eff}}^{(u)} &=& C_1 ({\cal{O}}_1^c - {\cal{O}}_1^u) + C_2 ({\cal{O}}_2^c - {\cal{O}}_2^u). 
 \end{eqnarray}
$C_{i} \equiv C_{i} (\mu)$ and ${\cal{O}}_{i} \equiv {\cal{O}}_{i} (\mu)$ are the 
Wilson coefficients and the local effective operators respectively. 
In order to study the most general NP in $b \to s \, l^+ l^-$ transitions, we augment ${\cal{H}}_{\mathrm{eff}}^{(t)}$ by 
\begin{eqnarray}
{\cal{H}}_{\mathrm{eff}}^{(t), \, \rm New} &=&  \sum_{i=7, 9, 10} C_{i^\prime} {\cal{O}}_{i^\prime} + 
\sum_{i=S, P} (C_i {\cal{O}}_i + C_{i^\prime} {\cal{O}}_{i^\prime}) + \sum_{i=T, T5} C_i {\cal{O}}_i \, , \nonumber \\
\end{eqnarray}
where the definitions of the local operators are given by,
\begin{multicols}{2}
\noindent
 \begin{align}
  \mathcal{O}_7     & = \frac{e}{16 \pi^2} m_b (\overline{s}\sigma_{\mu\nu} P_R b) F^{\mu\nu} \nonumber \\ 
  \mathcal{O}_9     & = \frac{\alpha_{\rm em}}{4 \pi} (\overline{s}\gamma_\mu P_L b) (\overline{l}\gamma^\mu l) \nonumber \\ 
  \mathcal{O}_{10}  & = \frac{\alpha_{\rm em}}{4 \pi} (\overline{s}\gamma_\mu P_L b) (\overline{l}\gamma^\mu \gamma_5 l) \nonumber \\ 
  \mathcal{O}_S     & = \frac{\alpha_{\rm em}}{4 \pi} (\overline{s} P_R b) (\overline{l} l) \nonumber \\ 
  \mathcal{O}_P     & = \frac{\alpha_{\rm em}}{4 \pi} (\overline{s} P_R b) (\overline{l} \gamma_5 l) \nonumber \\
  \mathcal{O}_T     & = \frac{\alpha_{\rm em}}{4 \pi} (\overline{s} \sigma_{\mu\nu} b) (\overline{l} \sigma^{\mu\nu} l) \nonumber
 \end{align}
 \begin{align}
   \mathcal{O}_{7^\prime}    & = \frac{e}{16 \pi^2} m_b (\overline{s}\sigma_{\mu\nu} P_L b) F^{\mu\nu} \nonumber \\ 
   \mathcal{O}_{9^\prime}    & = \frac{\alpha_{\rm em}}{4 \pi} (\overline{s}\gamma_\mu P_R b) (\overline{l}\gamma^\mu l) \nonumber \\ 
   \mathcal{O}_{10^\prime} & = \frac{\alpha_{\rm em}}{4 \pi} (\overline{s}\gamma_\mu P_R b) (\overline{l}\gamma^\mu \gamma_5 l) \nonumber \\ 
   \mathcal{O}_{S^\prime}    & = \frac{\alpha_{\rm em}}{4 \pi} (\overline{s} P_L b) (\overline{l} l) \nonumber \\ 
   \mathcal{O}_{P^\prime}    & = \frac{\alpha_{\rm em}}{4 \pi} (\overline{s} P_L b) (\overline{l} \gamma_5 l) \nonumber \\
   \mathcal{O}_{T5}  & = \frac{\alpha_{\rm em}}{4 \pi} (\overline{s} \sigma_{\mu\nu} b) (\overline{l} \sigma^{\mu\nu} \gamma_5 l) \nonumber 
\end{align}
\end{multicols}

where $P_{L,R}=(1 \mp \gamma_5)/2$ and $m_b \equiv m_b(\mu)$ denotes the running $b$ quark mass in the 
$\overline{\mathrm{MS}}$ scheme. 

Since $C_7$ and $C_9$ always appear in particular combinations with other $C_{i \leq 6}$ (the operators ${\cal O}_{i \leq 6}$ are identical 
to the $P_{i \leq 6}$ given in \cite{Chetyrkin:1996vx,Bobeth:1999mk}) in matrix elements, it is customary 
to define the following effective Wilson coefficients \cite{Chetyrkin:1996vx,Bobeth:1999mk},
\bal
C_7^{\rm eff}(\mu)  & =   C_7(\mu) - \frac{1}{3} C_3(\mu) -  \frac{4}{9} C_4(\mu) - \frac{20}{3} C_5(\mu) - \frac{80}{9} C_6(\mu) \, , \\
C_9^{\rm eff}(\mu) & =  C_9(\mu) + Y(q^2,\mu) \, ,
\eal
where the one loop expression for the function $Y(q^2,\mu)$ can be found in \cite{Bobeth:1999mk,Altmannshofer:2008dz}.

Note that, the photonic dipole operators  $\mathcal{O}_{7}$ and $\mathcal{O}_{7^\prime}$ lead to lepton universal contributions modulo 
lepton mass effects and hence, can not provide an explanation of the $R_{K^*}$ anomalies once bound from $B \to X_s \gamma$ is 
taken into account \cite{DescotesGenon:2011yn}. Moreover, as the tensor operators do not get 
generated at the dimension 6 level if the full SM gauge invariance is imposed \cite{Grzadkowski:2010es,Alonso:2014csa}, we ignore 
them in this work. 

\section{Results}
\label{results}

As the branching ratio of the fully leptonic decay $B_s \to \mu^+ \, \mu^-$ poses strong constraints on some of the Wilson coefficients, 
we first show the expression of this branching ratio as a function of the relevant couplings \cite{Alok:2010zd}, 
\begin{eqnarray}
{\cal B}(B_s \to \mu^+ \, \mu^-) & =& \frac{G^2_F \alpha_{em}^2
m^5_{B_s} f_{B_s}^2 \tau_{B_s}}{64 \pi^3}
|V_{tb}^{}V_{ts}^{\ast}|^2 \sqrt{1 - \frac{4 m_\mu^2}{m_{B_s}^2}}\times
\nn\\
&& \hskip-2truecm  \Bigg\{
\Bigg(1 - \frac{4m_\mu^2}{m_{B_s}^2} \Bigg) \Bigg|
\frac{C_S^\mu - C_{S'}^\mu}{m_b + m_s}\Bigg|^2
+ \Bigg|\frac{C_P^\mu - C_{P'}^\mu}{m_b + m_s} + \frac{2 m_\mu}{m^2_{B_s}} (C_{10}^{\rm SM}
+\Delta C_{10}^\mu - C_{10'}^\mu)\Bigg|^2 \Bigg\}. \phantom{space}
\label{bmumu-BR}
\end{eqnarray}

\begin{figure}[ht!]
\begin{center}
\begin{tabular}{cc}
\includegraphics[scale=0.5]{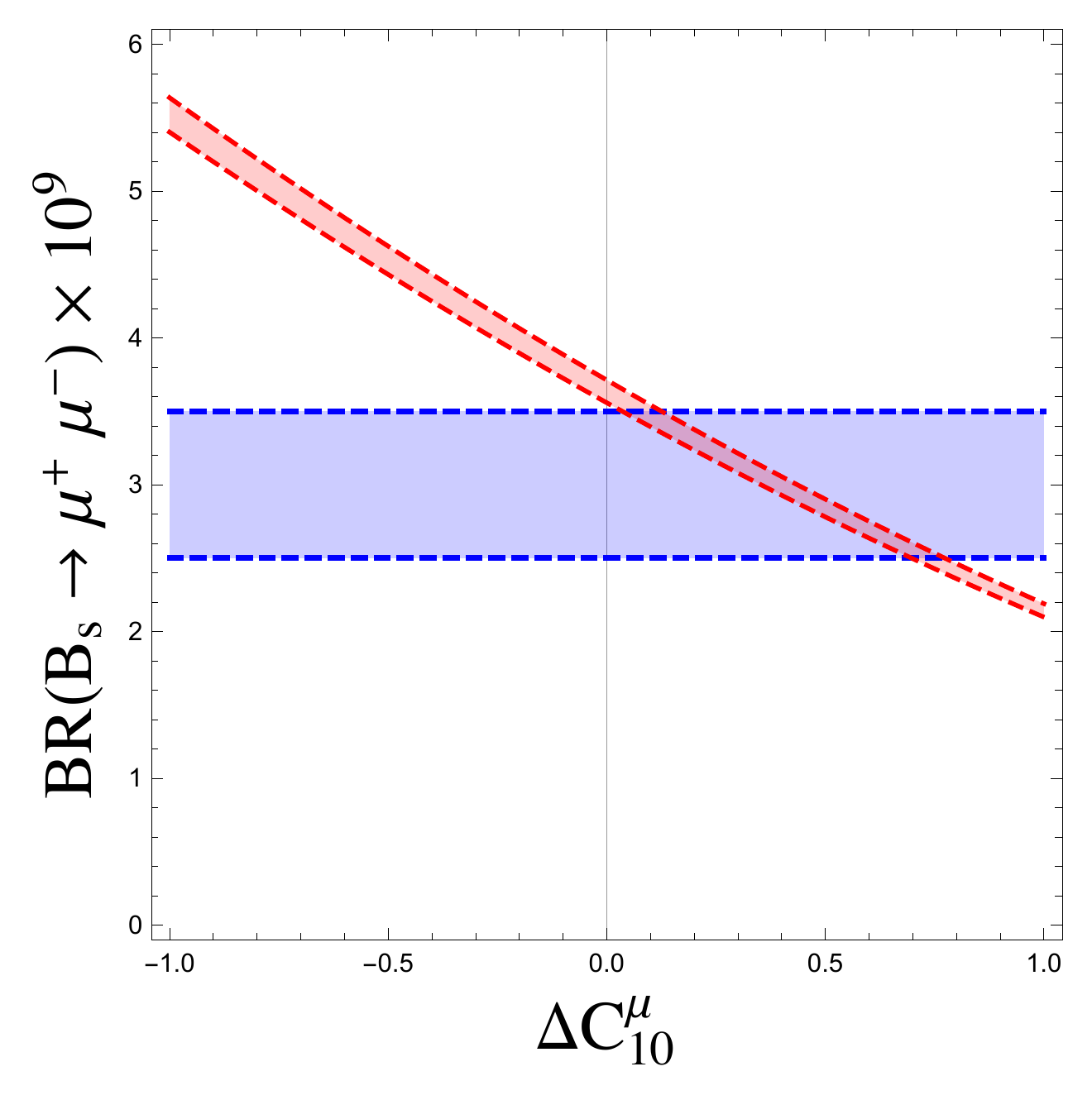} & \includegraphics[scale=0.5]{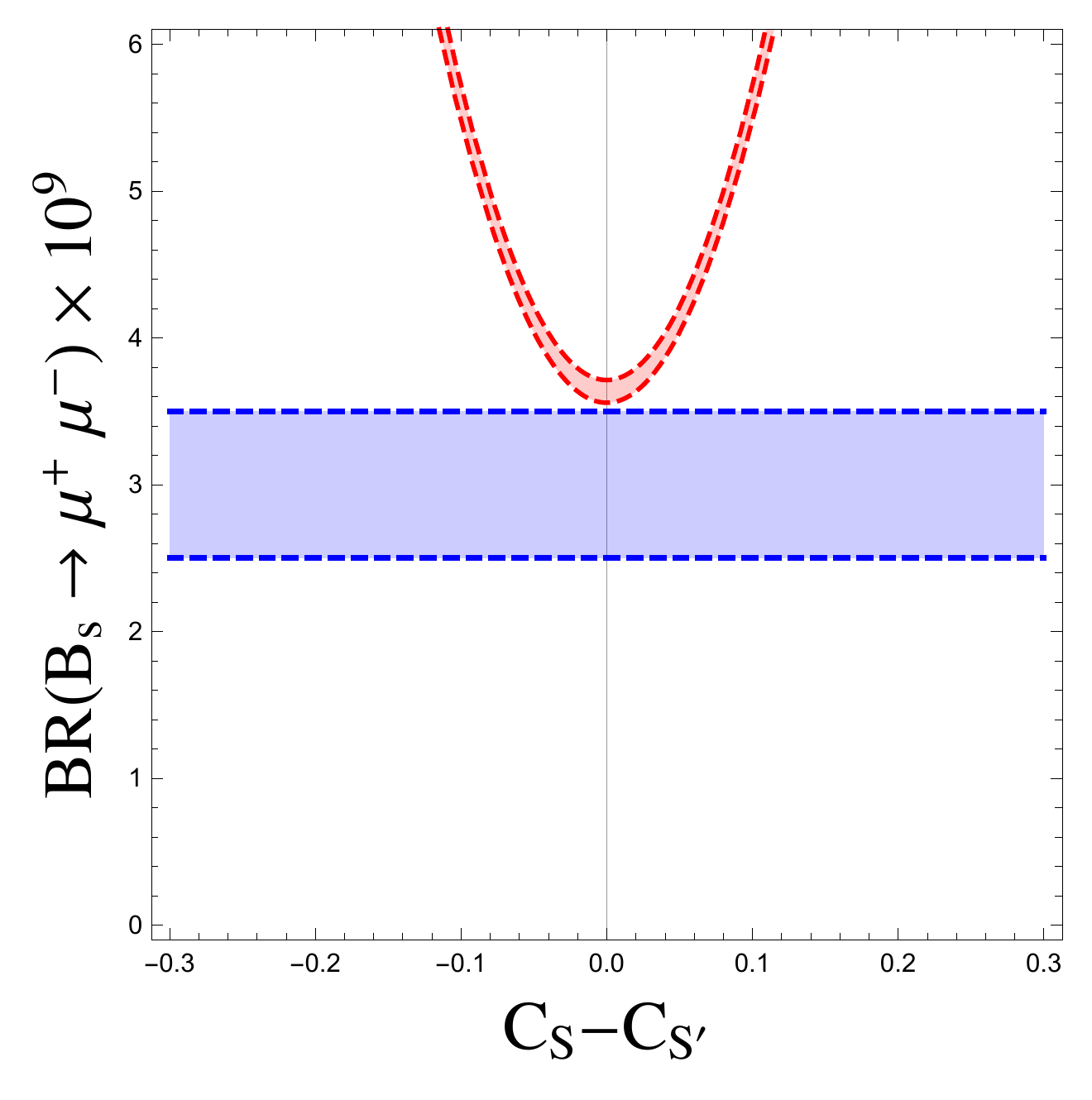} \\
\includegraphics[scale=0.5]{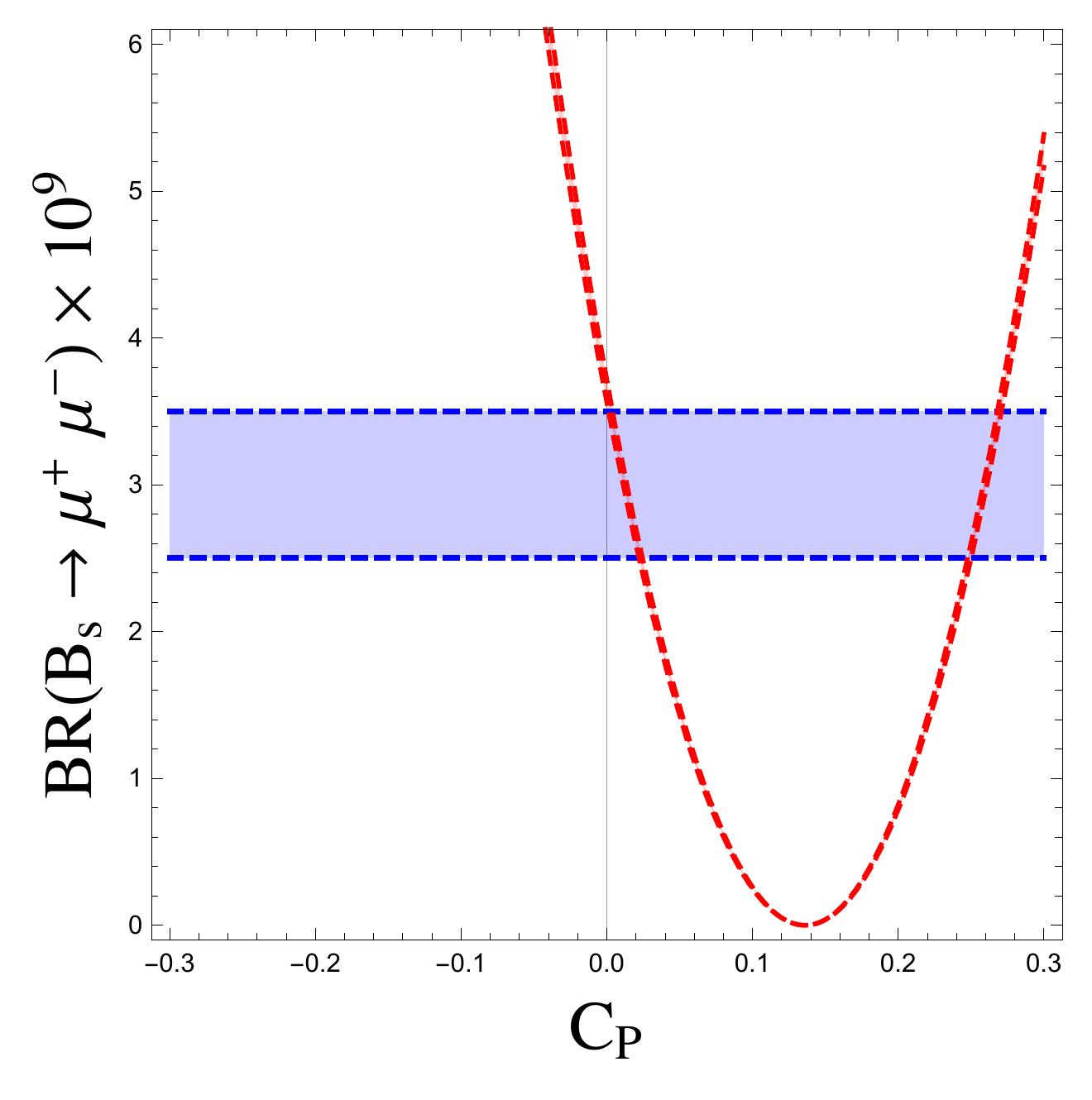} & \includegraphics[scale=0.5]{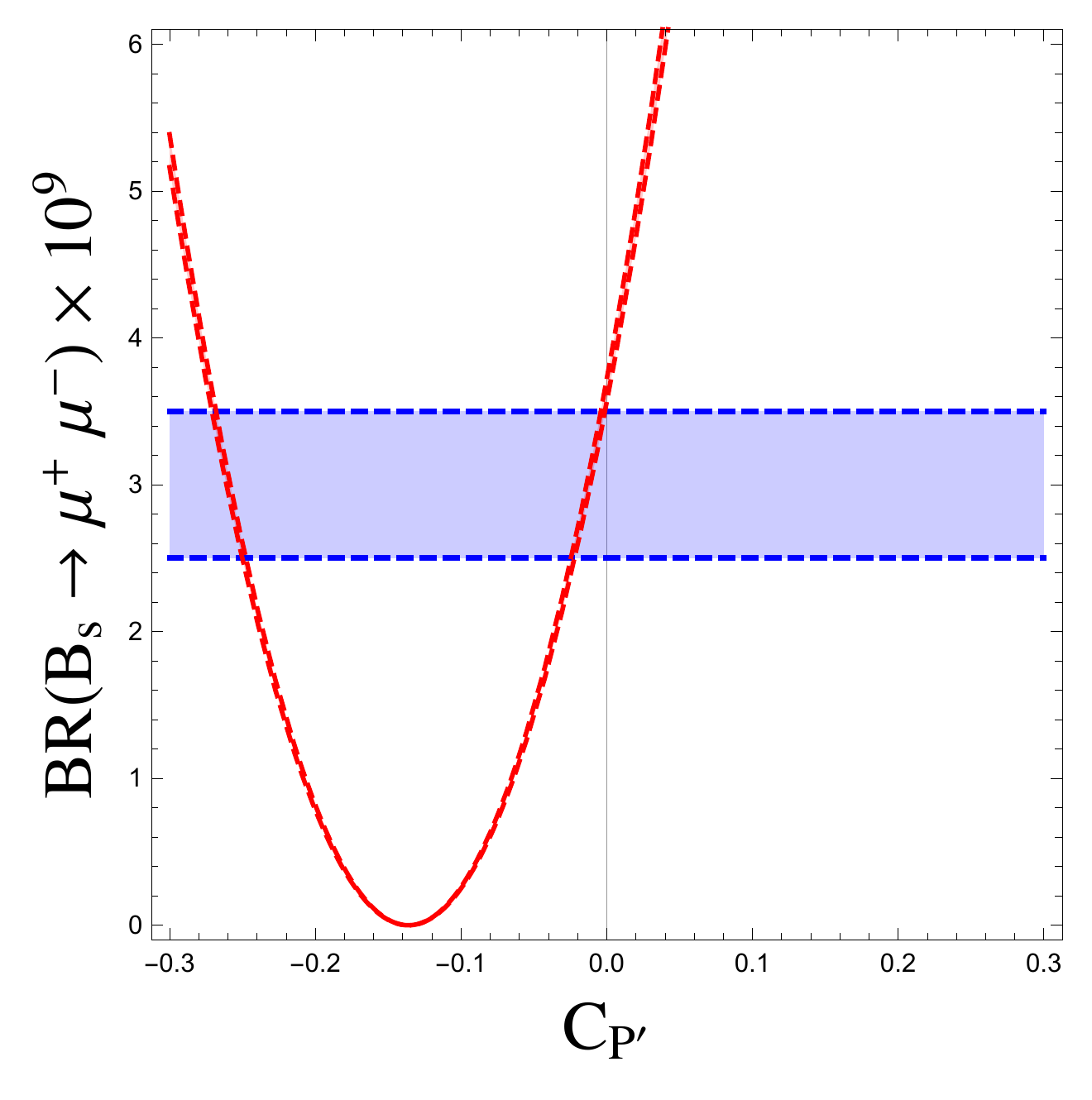} 
\end{tabular}
\caption{Variation of the branching ratio of $B_s \to \mu^+ \, \mu^-$ with $\Delta C_{10}^\mu$ and the (pseudo) scalar operaors.\label{Bsmumu}}
\end{center}
\end{figure}

In Fig.~\ref{Bsmumu}, we show how ${\cal B}(B_s \to \mu^+ \, \mu^-)$ constraints $\Delta C_{10}^\mu$, and also the scalar and pseudo scalar 
operators. The horizontal blue band shows the $1\sigma$ experimentally allowed region. Hence, 
$\Delta C_{10}^\mu$ ($\Delta C_{10'}^\mu$) should satisfy $0 \lesssim \Delta C_{10}^\mu \lesssim 0.7$ ($-0.7 \lesssim \Delta C_{10'}^\mu \lesssim 0$).  

Note that, unlike $\Delta C_{10}^\mu$ and $\Delta C_{10'}^\mu$, there are practically no bounds on 
$\Delta C_{10}^e$ and $\Delta C_{10'}^e$ because the experimental upper bound, $2.8 \times 10^{-7}$ \cite{Aaltonen:2009vr}, 
is many orders of magnitude above the SM prediction $(8.35 \pm 0.39) \times 10^{-14}$ \cite{Bobeth:2013uxa,Fleischer:2017ltw}. 

The constraints on the scalar operators are extremely severe, as can be seen from the figures. 

\subsection{One Wilson coefficient at a time}
In this section, we consider one Wilson coefficient at a time and investigate whether it can explain all the experimental results 
within their $1\sigma$ values simultaneously. All the numerical results in this section are based on the analytic formulas given 
in \cite{Alok:2009tz, Alok:2010zd}. As for the form-factors, we use \cite{Bouchard:2013pna} for $B \to K$ matrix elements and 
\cite{Straub:2015ica} for the $B \to K^*$ matrix elements.

\subsubsection*{Scalar and pseudo scalar operators:}

\begin{figure}[!ht]
\begin{center}
\begin{tabular}{cc}
\includegraphics[scale=0.8]{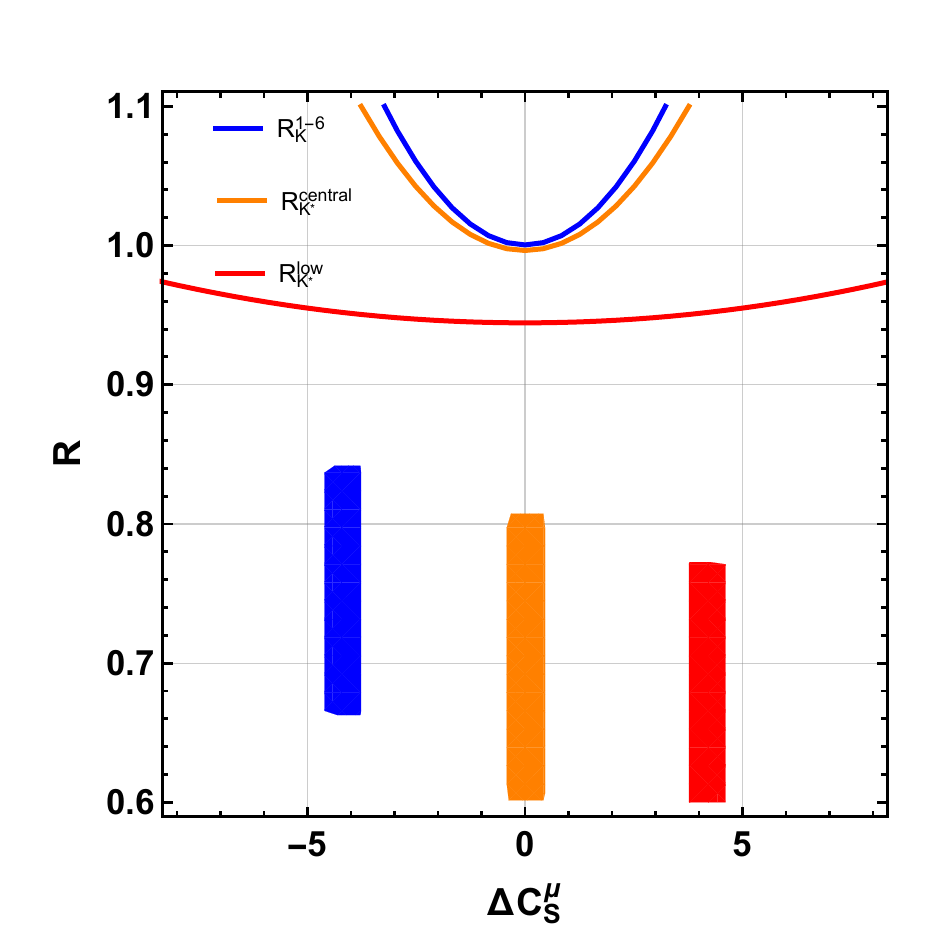} & \includegraphics[scale=0.8]{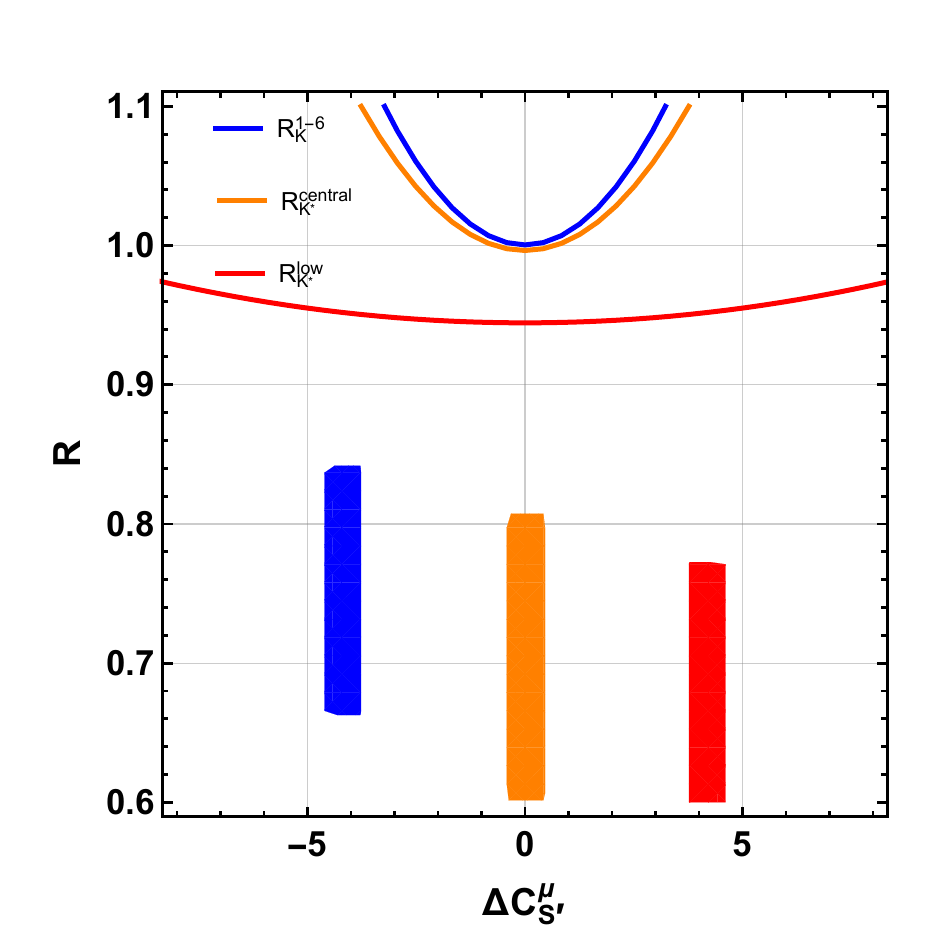} \\
\includegraphics[scale=0.8]{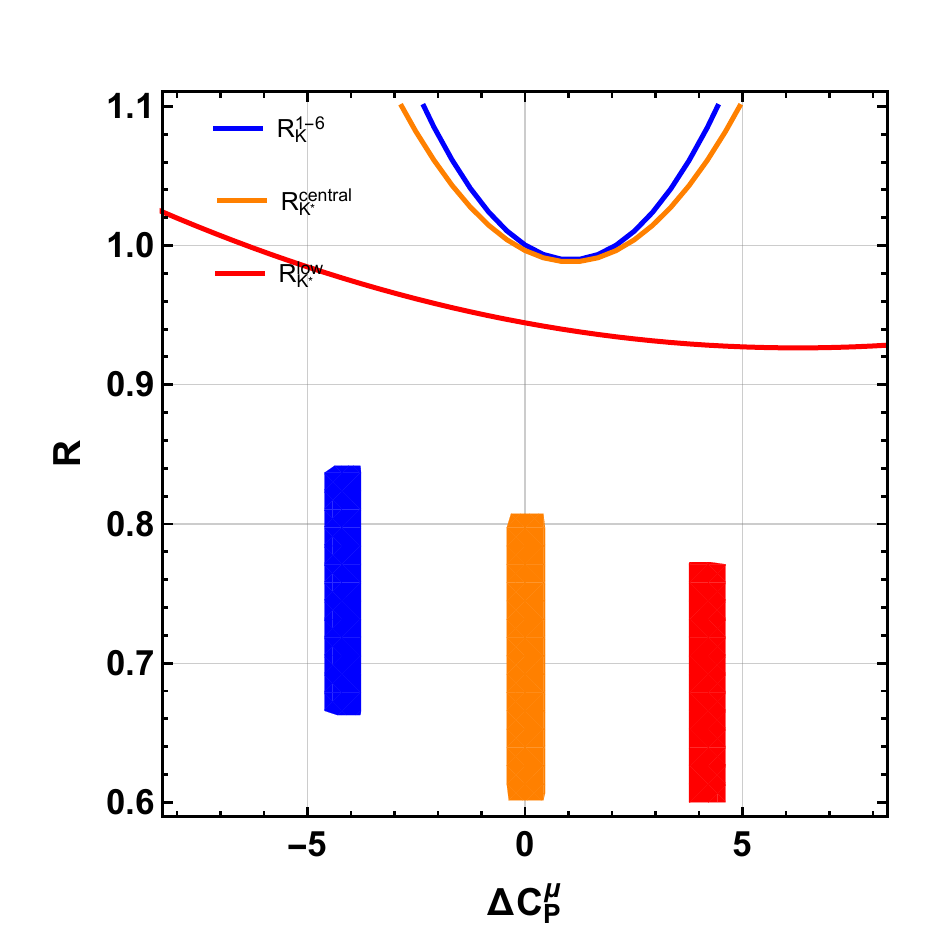} & \includegraphics[scale=0.8]{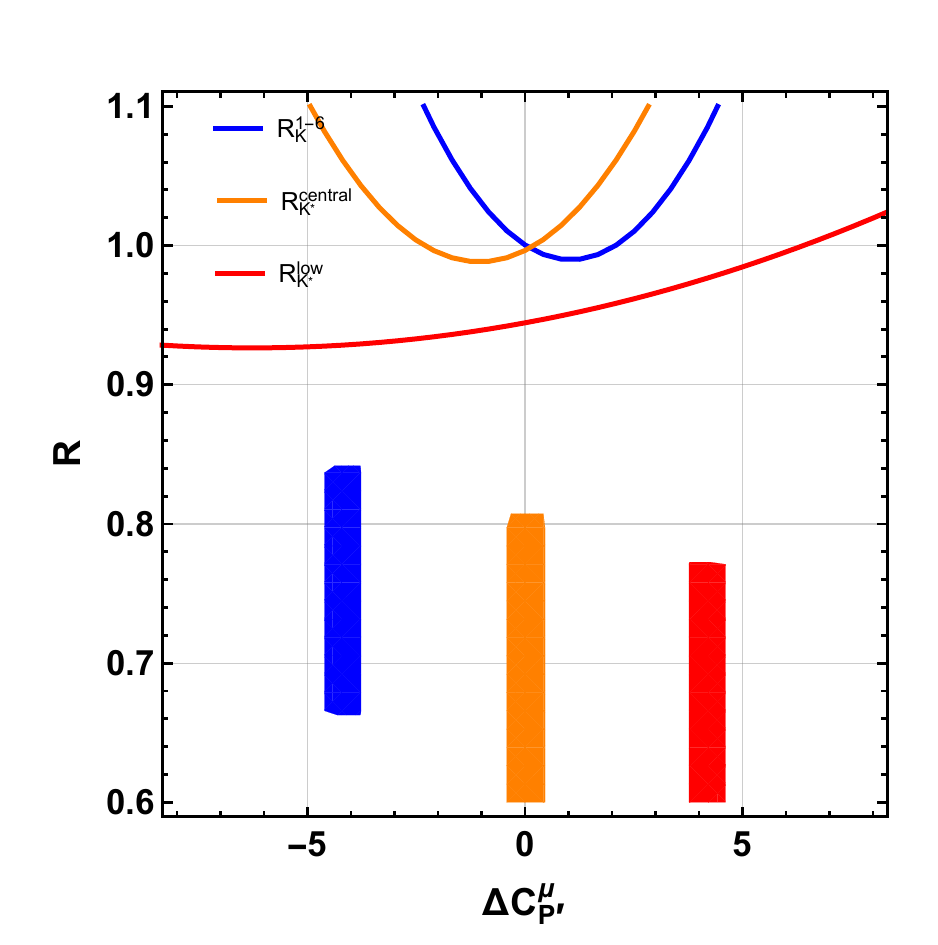}
\end{tabular}
\caption{Variations of $R_{K}$, $R_{K^*}^{\rm low}$ and $R_{K^*}^{\rm central}$ with the Wilson coefficients of the various scalar 
and pseudo-scalar operators involving muons. The vertical bands correspond to the experimental $1 \sigma$ allowed regions 
(and independent of $\Delta C$). \label{scalar-mu}}
\end{center}
\end{figure}

We first present our results for the scalar operators. In Fig.~\ref{scalar-mu} we show  $R_{K}$, $R_{K^*}^{\rm low}$ and 
$R_{K^*}^{\rm central}$ as functions of the scalar and pseudo-scalar Wilson coefficients $C_S$, $C_{S^\prime}$, $C_{P}$ and  
$C_{P^\prime}$ assuming that they only affect the muon mode. It is clear from the plots that 
(pseudo) scalar operators involving muons are unable to provide solutions to these anomalies, irrespective of their size. 

\begin{figure}[!ht]
\begin{center}
\begin{tabular}{cc}
\includegraphics[scale=0.8]{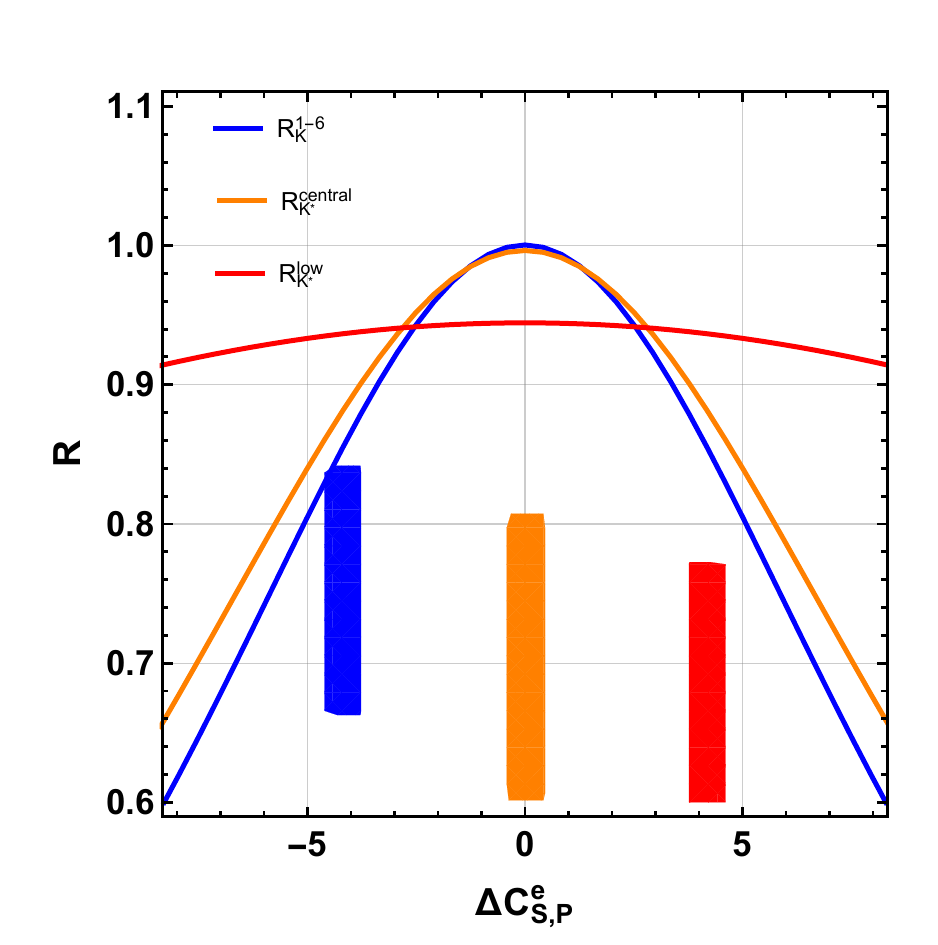} & \includegraphics[scale=0.8]{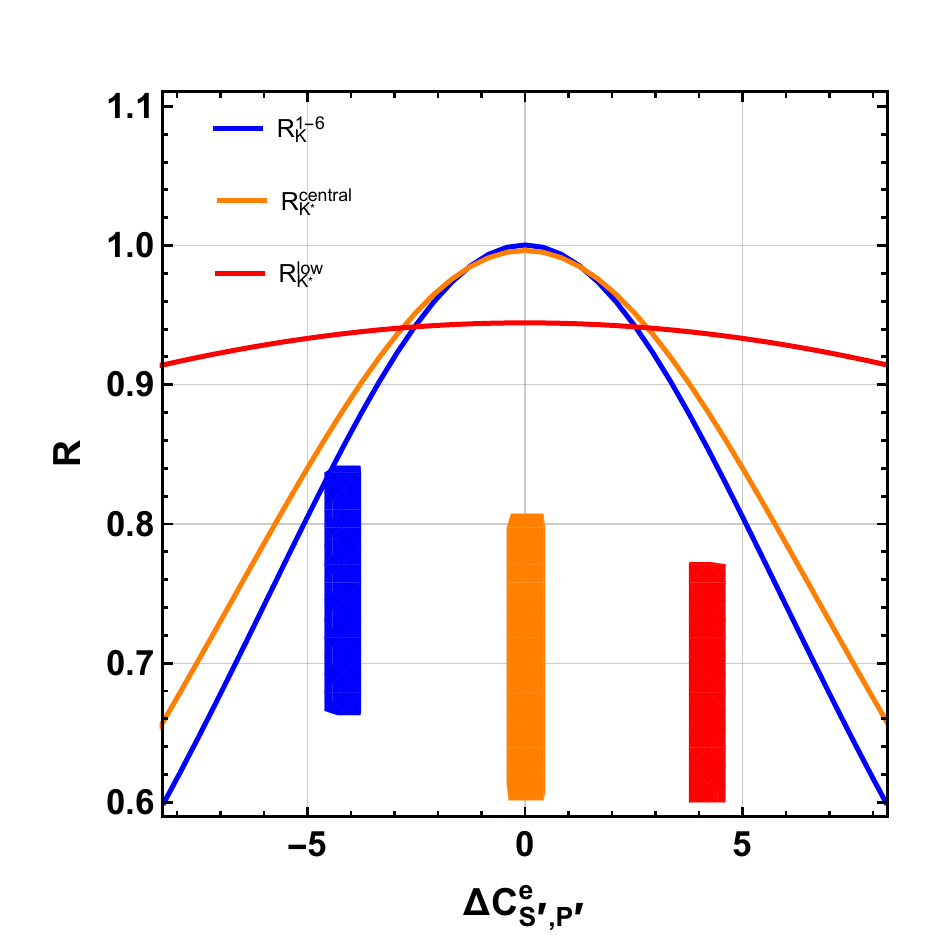}
\end{tabular}
\caption{Variations of $R_{K}$, $R_{K^*}^{\rm low}$ and $R_{K^*}^{\rm central}$ with the Wilson coefficients of the various scalar 
and pseudo-scalar operators involving electrons. 
The variations with the different Wilson coefficients are the same in this case because the decay rate for $B \to K^{(*)} e^+ e^- $ dominantly depends 
on their modulus squared with same coefficients. The linear interference terms, which have different coefficients for the different operators, 
are negligible because they are proportional to the electron mass \cite{Alok:2010zd}. The vertical bands correspond to the experimental $1 \sigma$ 
allowed regions (and independent of $\Delta C$).\label{scalar-e}}
\end{center}
\end{figure}

It can be seen form Fig.~\ref{scalar-e} that the 
same statement is also true for the (pseudo) scalar operators involving electrons. However, for the operators involving electrons, solutions 
to two of the anomalies, $R_{K}$ and $R_{K^*}^{\rm central}$, are in principle possible. But, the upper bound on 
$\mathcal{B} \left( B_s \to e^+ e^-\right)$ (see table~\ref{exp-data}) constrains the couplings $C_{S,S',P,P'} \lesssim 1.2 $, and rules out 
the possibility of any such explanations.

\subsubsection*{Vector and axial vector operators:}

\begin{figure}[ht!]
\begin{center}
\begin{tabular}{cc}
\hspace*{-1cm}\includegraphics[scale=0.85]{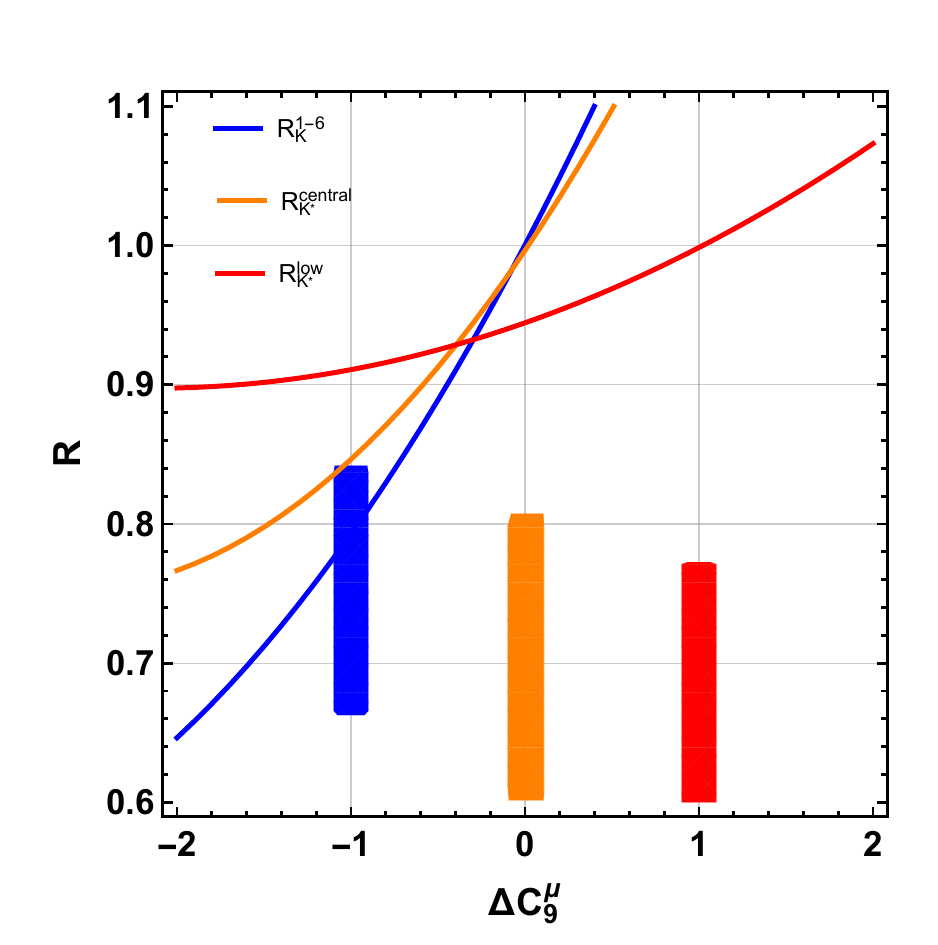} & \includegraphics[scale=0.85]{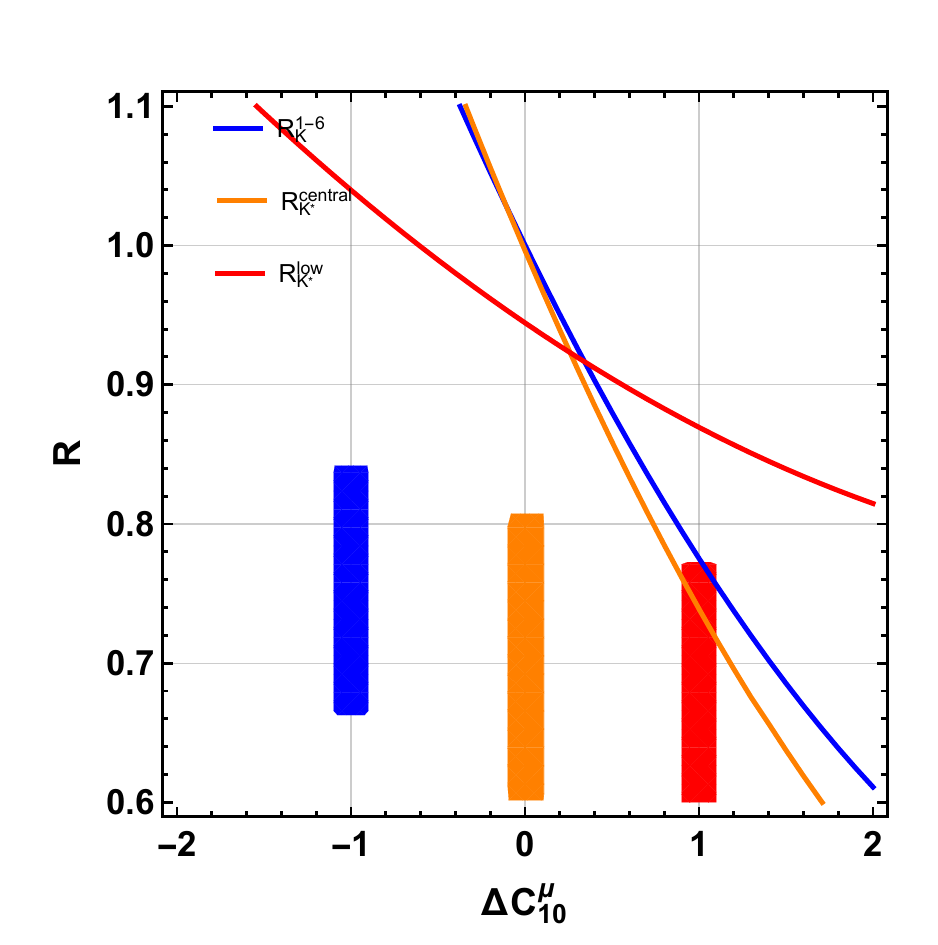} \\
\hspace*{-1cm}\includegraphics[scale=0.85]{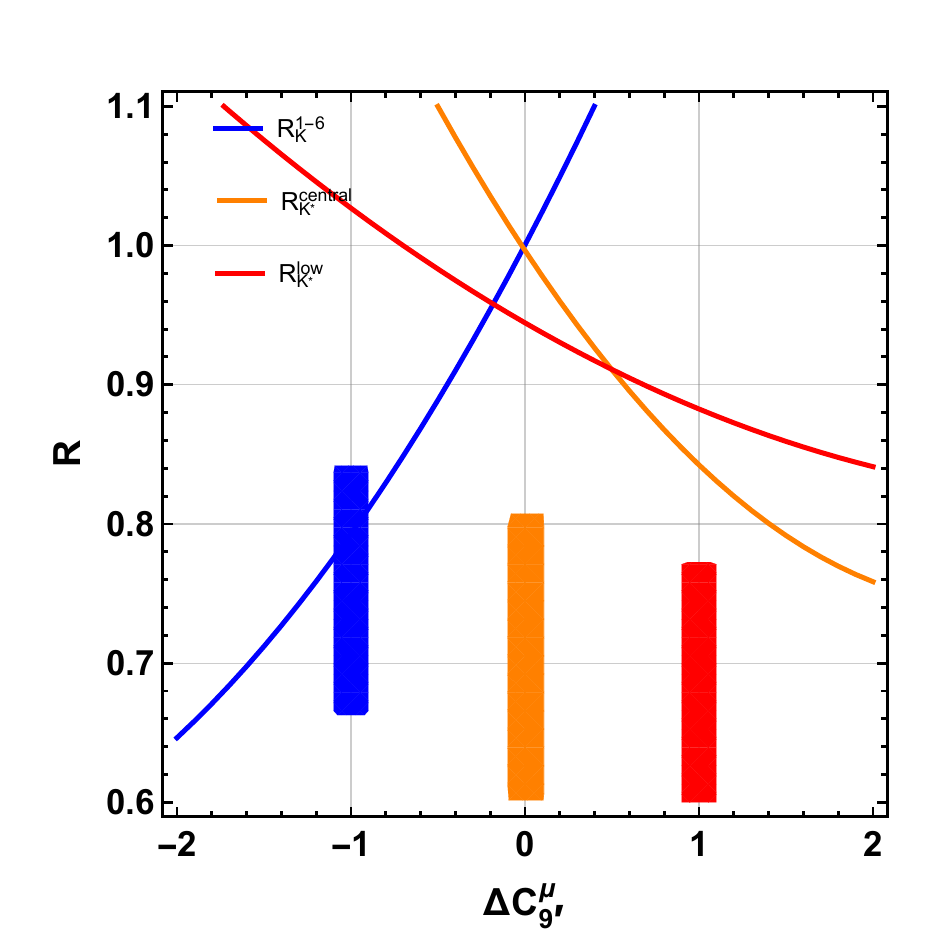} & \includegraphics[scale=0.85]{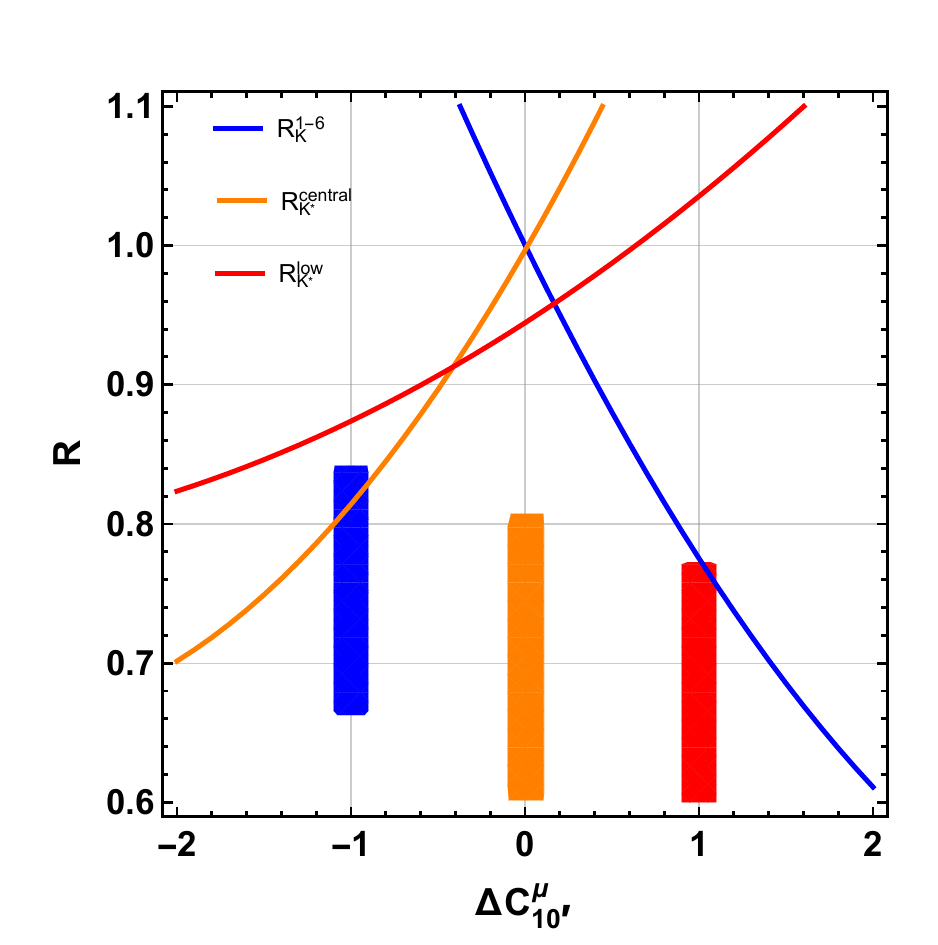}
\end{tabular}
\caption{Variations of $R_{K}$, $R_{K^*}^{\rm low}$ and $R_{K^*}^{\rm central}$ with the various vector and axial vector 
Wilson coefficients in the muon mode. The vertical bands correspond to the experimental $1 \sigma$ allowed regions 
(and independent of $\Delta C$). The legends explain the meaning of the different colours. 
We only plot the central values of the observables as the uncertainties are expected to be very 
small in these ratios, see for example \cite{Capdevila:2017bsm,Altmannshofer:2017yso,Geng:2017svp}. \label{vector-mu}}
\end{center}
\end{figure}

We now turn to the vector and axial vector operators. 
Fig.~\ref{vector-mu} shows the variations of $R_{K}$, $R_{K^*}^{\rm low}$ and $R_{K^*}^{\rm central}$ with respect to the Wilson coefficients   
$C_9^\mu$, $C_{9^\prime}^\mu$, $C_{10}^\mu$ and  $C_{10^\prime}^\mu$. 
It can be seen that even the vector and axial vector operators in the muon mode, when taken one at a time,  can not explain all 
the anomalies within their experimental $1 \sigma$ regions.  Additionally, as mentioned after Eq.~\ref{bmumu-BR}, the axial vector operators 
$ \Delta C_{10}^\mu$ and $\Delta C_{10'}^\mu$ are constrained rather strongly by measurement of the branching ratio of 
$B_s \to \mu^+ \, \mu^-$ : $0 \lesssim \Delta C_{10}^\mu \lesssim 0.7$  and $-0.7 \lesssim \Delta C_{10'}^\mu \lesssim 0$. 
This makes the axial-vector explanation even more unlikely.

\begin{figure}[ht!]
\begin{center}
\begin{tabular}{cc}
\hspace*{-1cm}\includegraphics[scale=0.85]{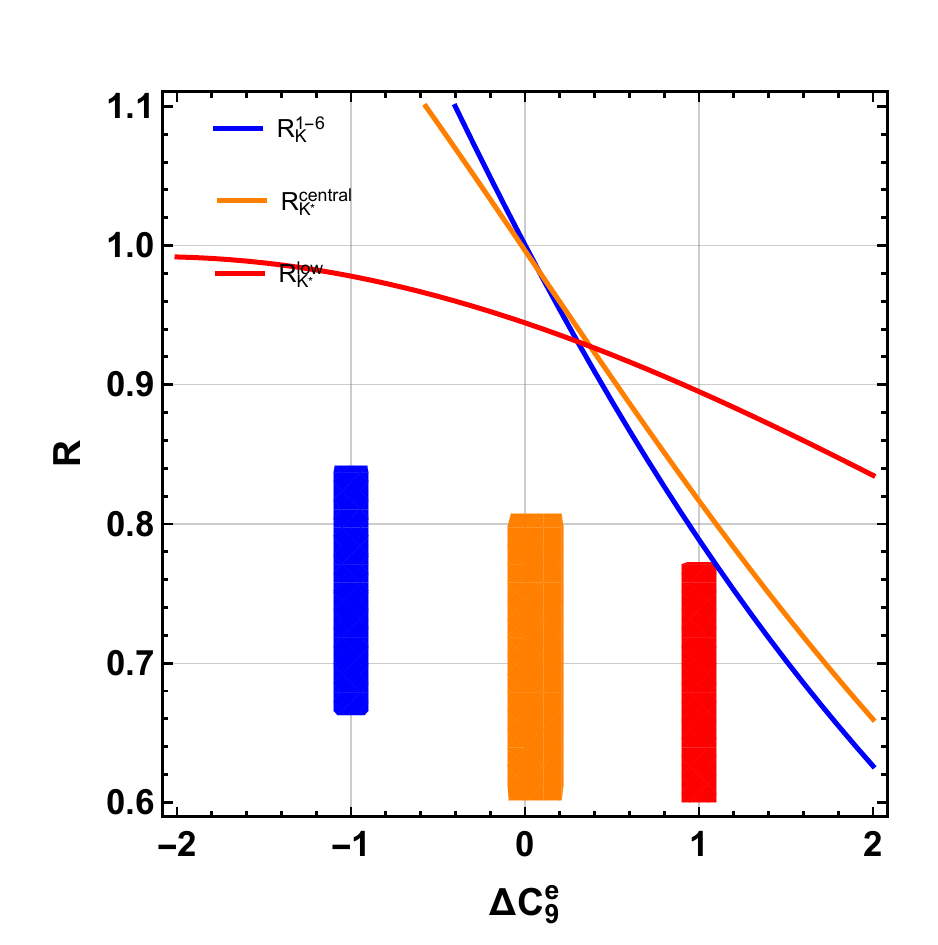} & \includegraphics[scale=0.85]{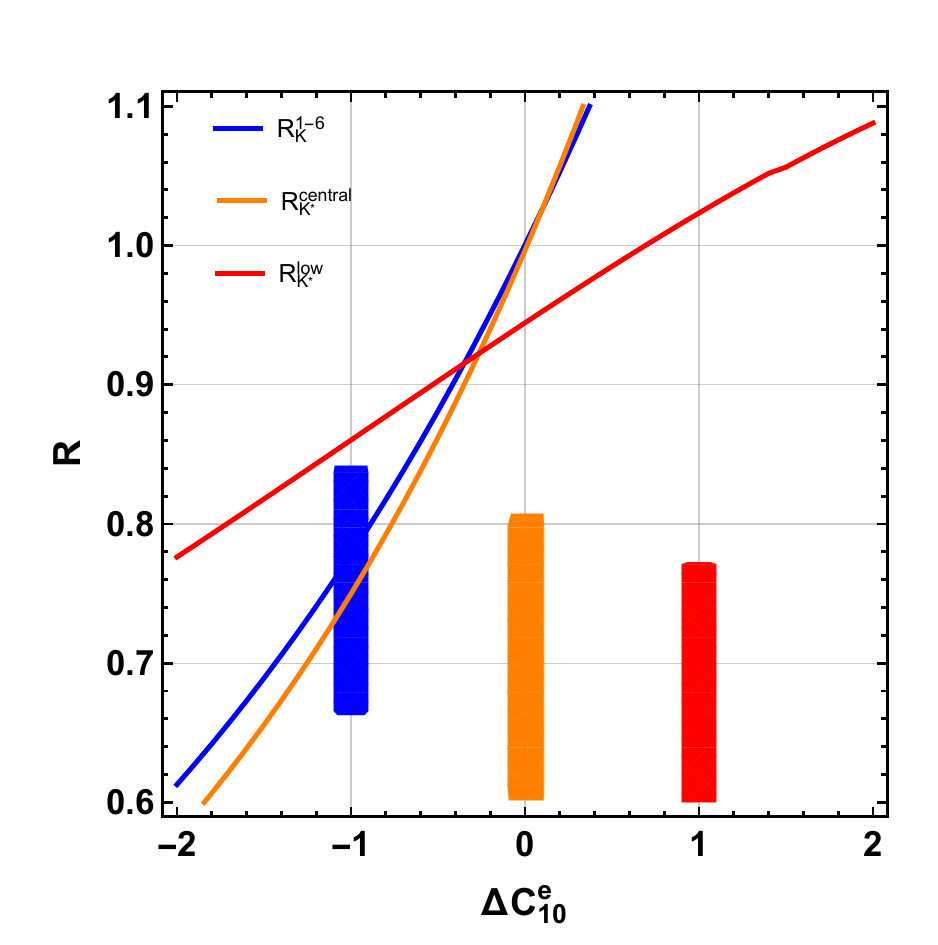} \\
\hspace*{-1cm}\includegraphics[scale=0.85]{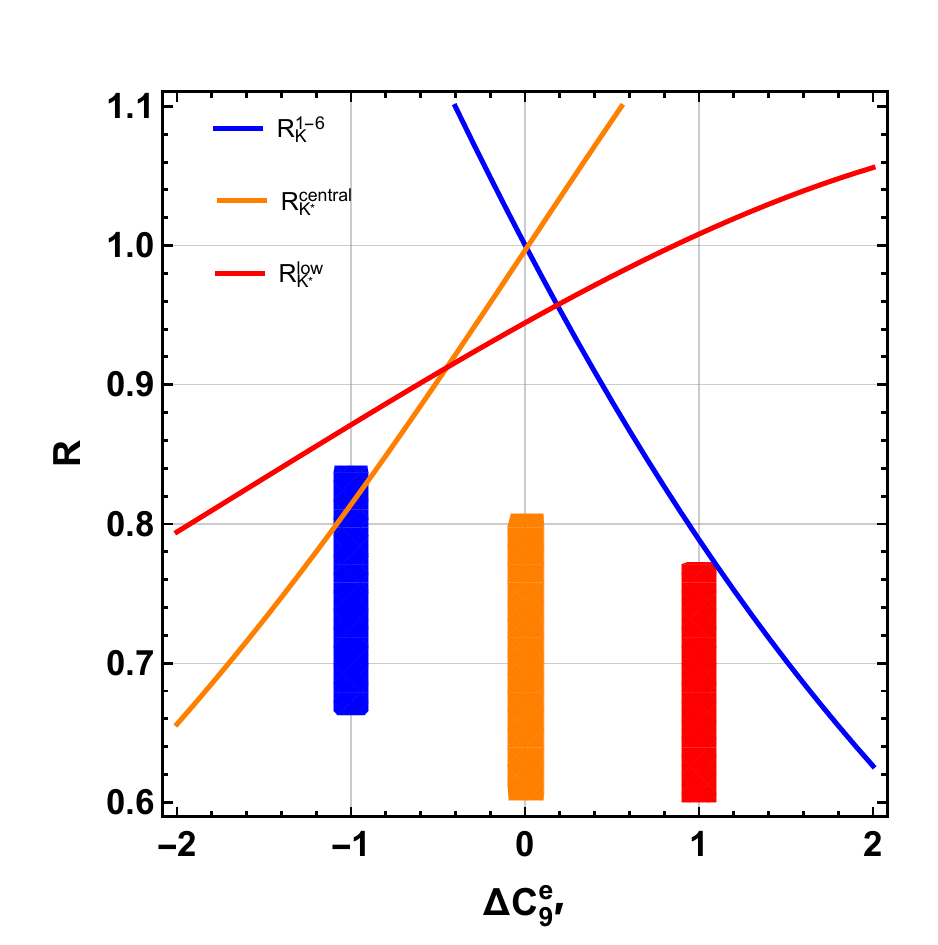} & \includegraphics[scale=0.85]{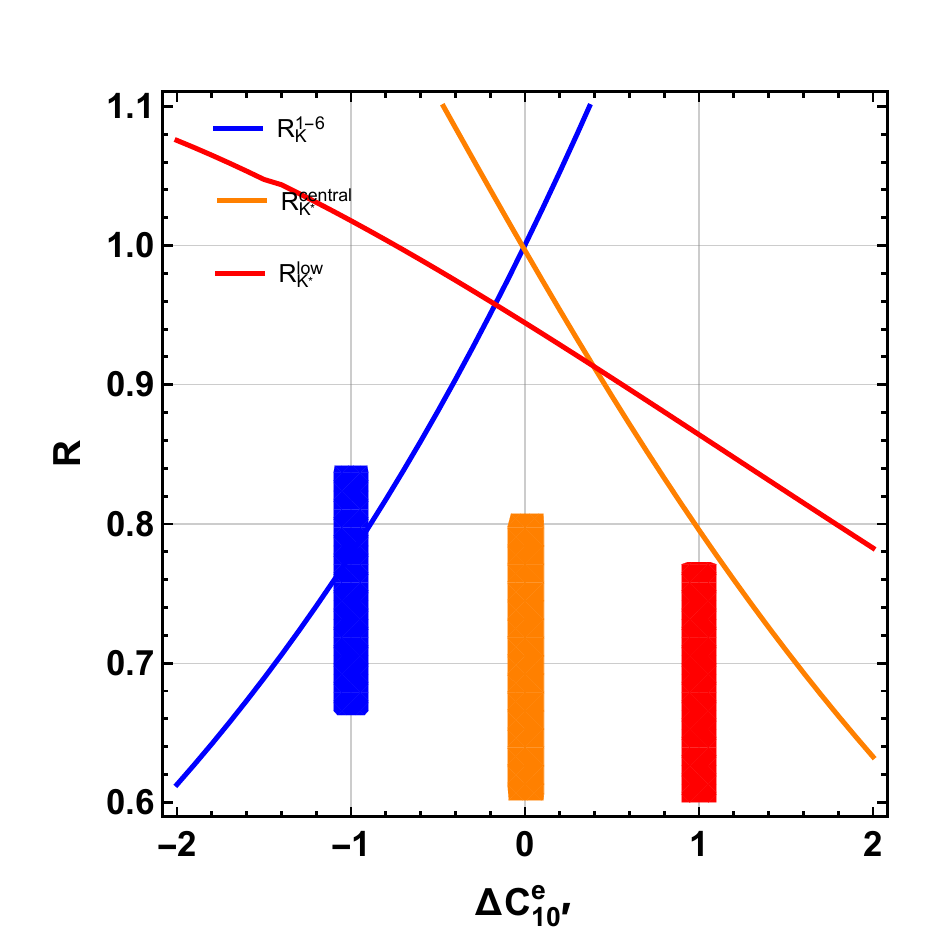}
\end{tabular}
\caption{Variations of $R_{K}$, $R_{K^*}^{\rm low}$ and $R_{K^*}^{\rm central}$ with the various vector and axial vector 
Wilson coefficients in the electron mode. The vertical bands correspond to the experimental $1 \sigma$ allowed regions 
(and independent of $\Delta C$). The legends explain the meaning of the different colours. 
We only plot the central values of the observables as the uncertainties are expected to be very 
small in these ratios, see for example \cite{Capdevila:2017bsm,Altmannshofer:2017yso,Geng:2017svp}. \label{vector-e}}
\end{center}
\end{figure}

Similar statement can also be made about the (axial) vector operators in the electron sector, as can be seen in Fig.~\ref{vector-e}. 
However, they do a better job compared to their counterparts in the muon sector. While the primed operators are strongly 
disfavoured, the operator $\Delta C_{10}^e$ does comparatively better. For example, $\Delta C_{10}^e = -1.5$ gives 
$R_K = 0.69, R_{K^*}^{\rm central} = 0.66, R_{K^*}^{\rm low} = 0.81$, the first two numbers being inside their experimental 
$1 \sigma$ regions, and the value of $R_{K^*}^{\rm low}$ is $\sim 1.4 \sigma$ away from the experimental central value.

\clearpage
\subsection{Combination of Wilson coefficients}

\begin{figure}[h!]
\begin{center}
\begin{tabular}{cc}
\includegraphics[scale=0.8]{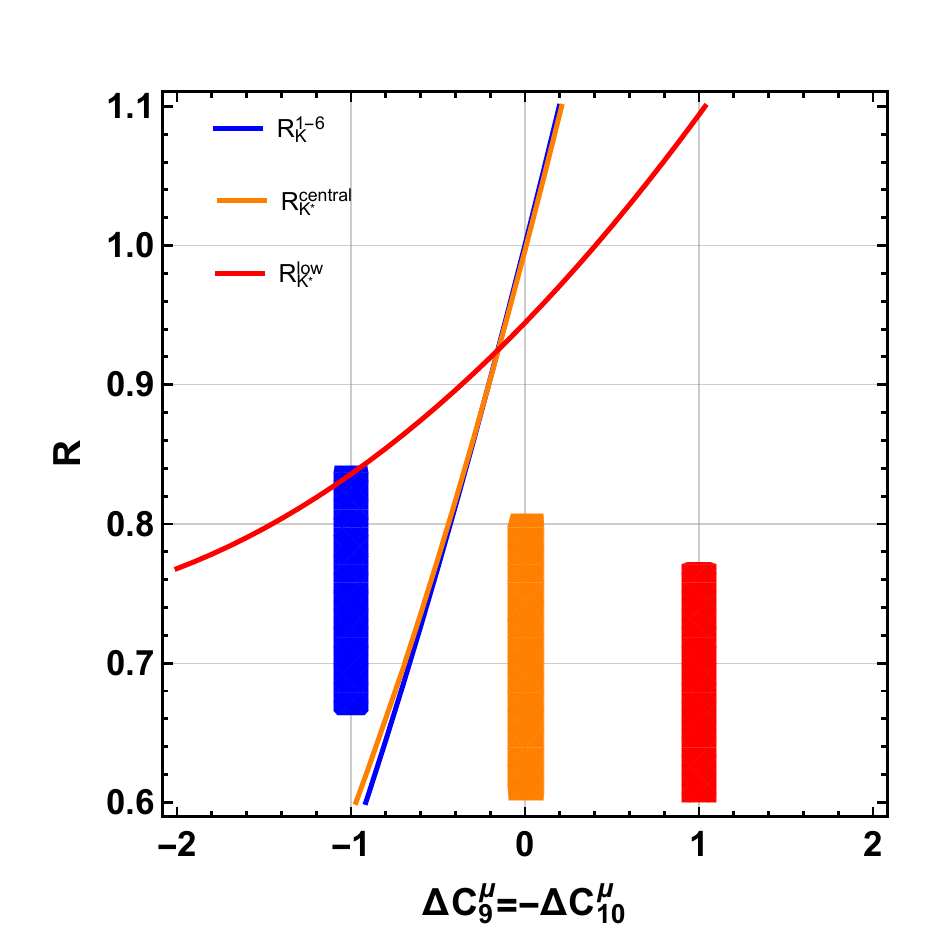} & \includegraphics[scale=0.8]{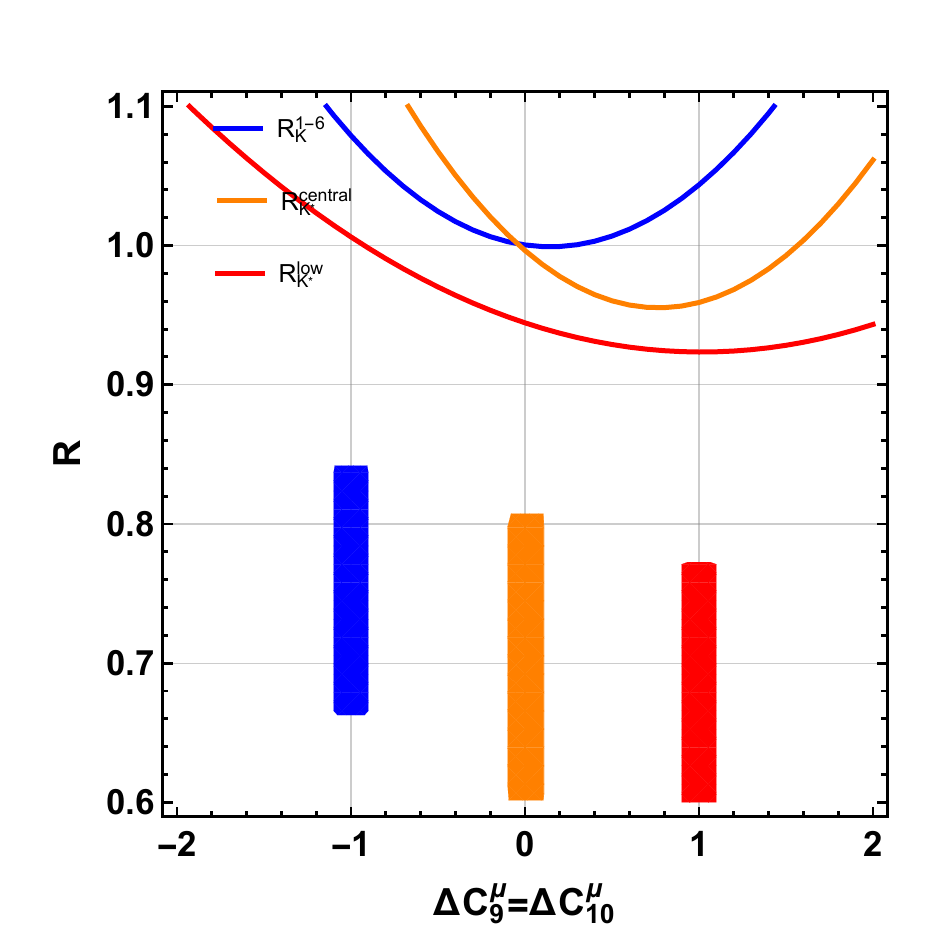} \\
\includegraphics[scale=0.8]{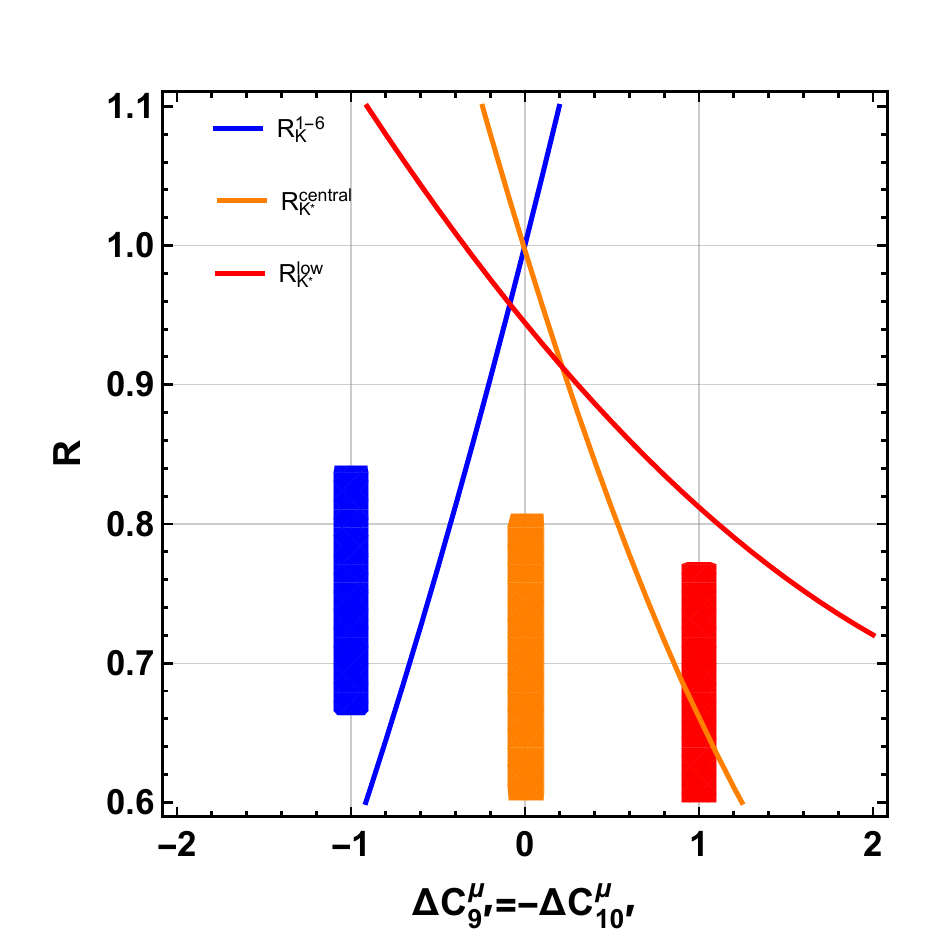} & \includegraphics[scale=0.8]{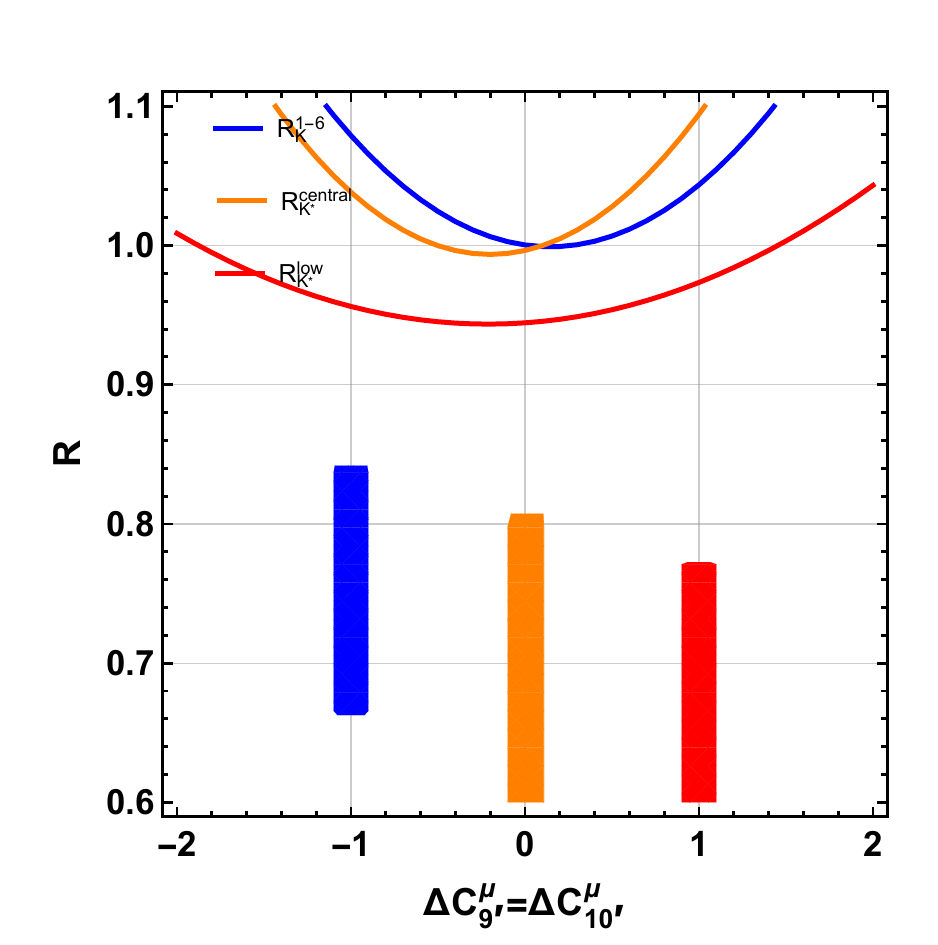}
\end{tabular}
\caption{Variations of $R_{K}$, $R_{K^*}^{\rm low}$ and $R_{K^*}^{\rm central}$ with the various vector and axial vector 
Wilson coefficients in the muon mode. The vertical bands correspond to the experimental $1 \sigma$ allowed regions (and independent of $\Delta C$). \label{comb-mu}}
\end{center}
\end{figure}

\begin{figure}[h!]
\begin{center}
\begin{tabular}{cc}
\includegraphics[scale=0.8]{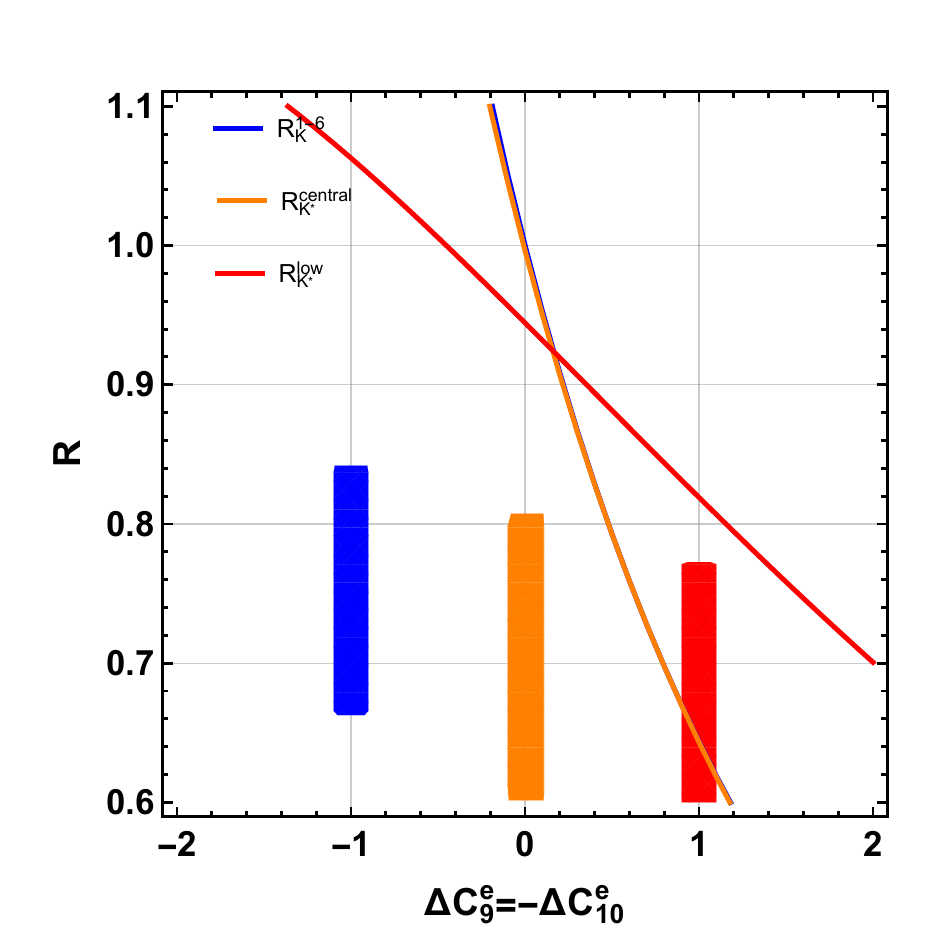} & \includegraphics[scale=0.8]{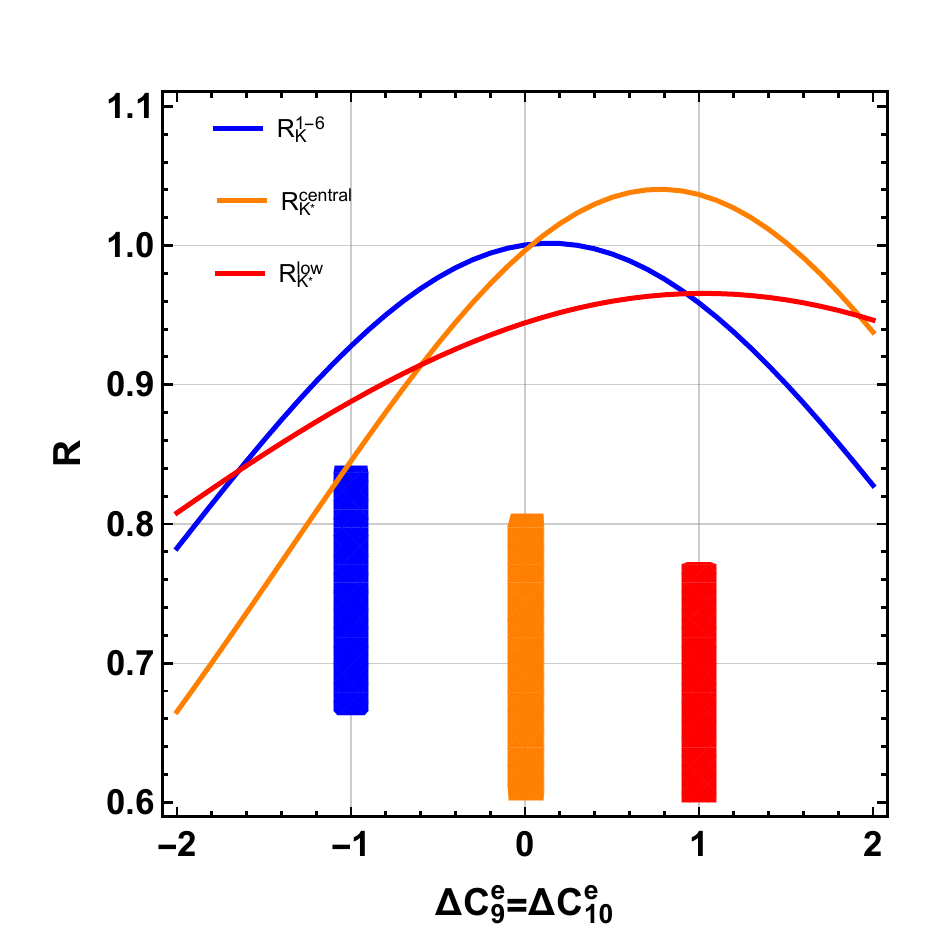} \\
\includegraphics[scale=0.8]{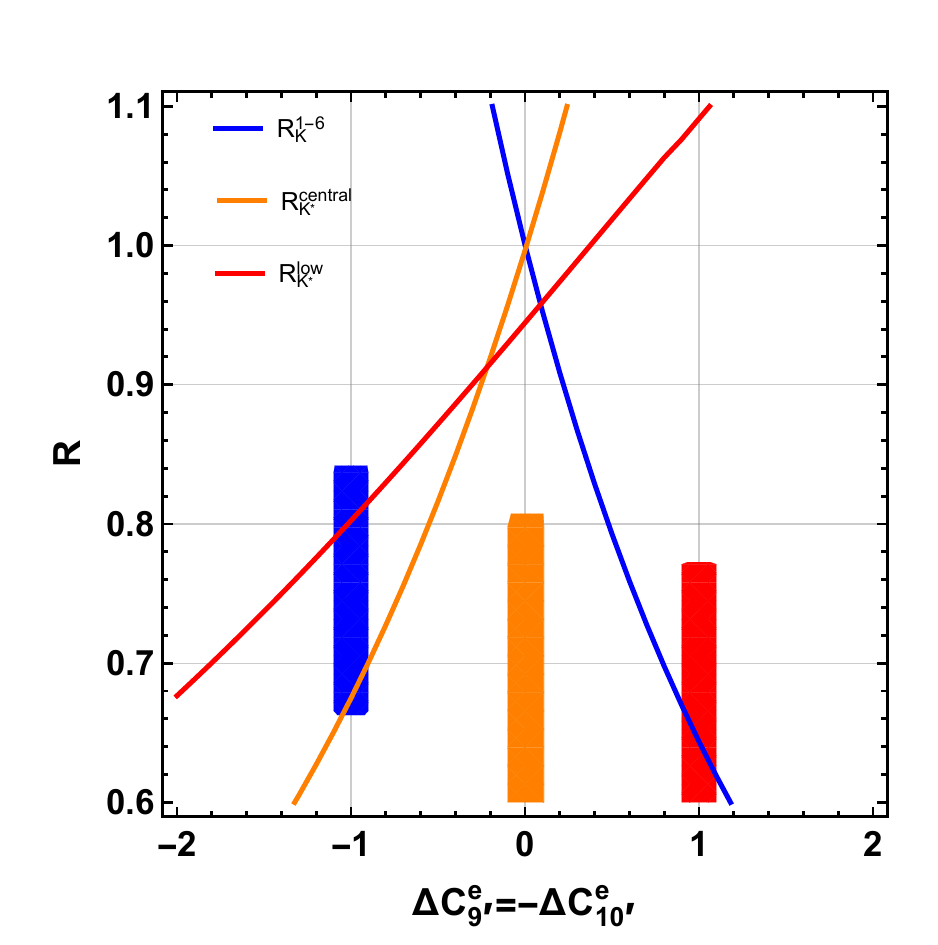} & \includegraphics[scale=0.8]{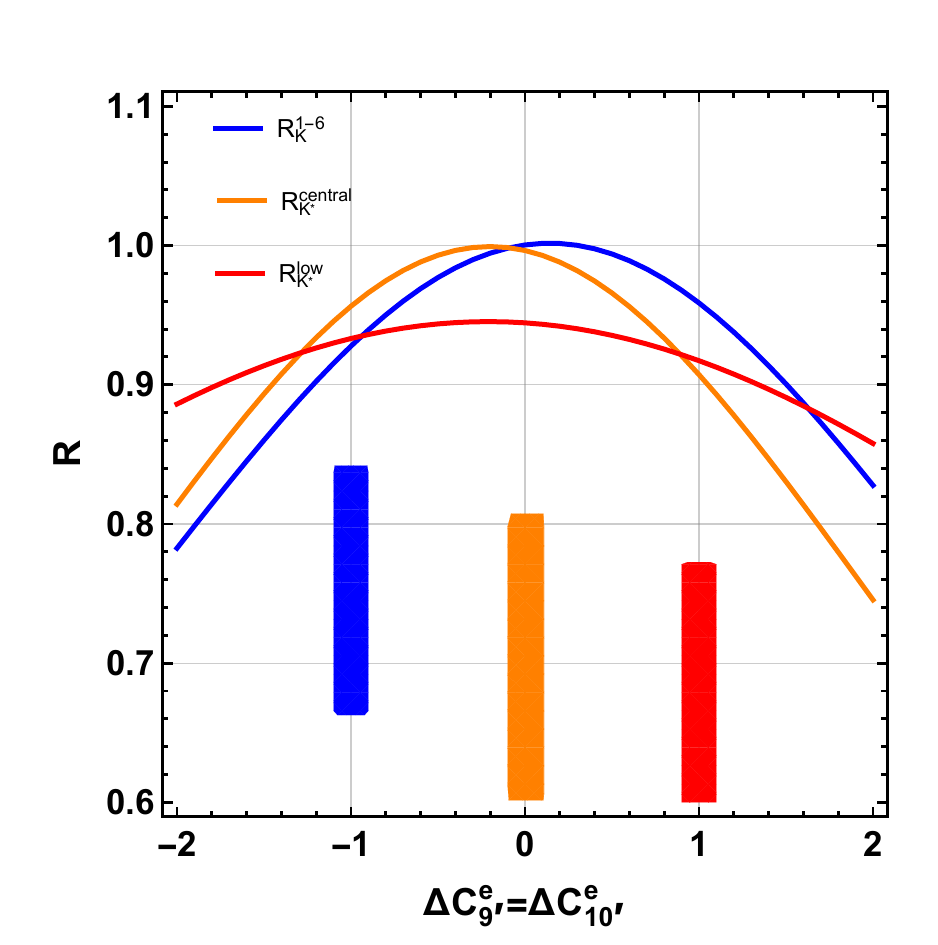}
\end{tabular}
\caption{Variations of $R_{K}$, $R_{K^*}^{\rm low}$ and $R_{K^*}^{\rm central}$ with the various vector and axial vector 
Wilson coefficients in the electron mode. The vertical bands correspond to the experimental $1 \sigma$ allowed regions (and independent of $\Delta C$). \label{comb-e}}
\end{center}
\end{figure}

In this section, we consider the four cases $\Delta C^{\ell}_{9^{(')}} = \pm \Delta C^{\ell}_{10^{(')}}$ for each of $\ell = \mu$ and e. 
The results are shown in Fig.~\ref{comb-mu} and \ref{comb-e} for Wilson coefficients involving muons and electrons respectively. 
The hypotheses $\Delta C^{\mu}_{9^{(')}} = \Delta C^{\mu}_{10^{(')}} $ (which correspond to the operators $(\bar s \gamma_\alpha 
P_L b ) (\bar \mu \gamma^\alpha P_R \mu)$ and $(\bar s \gamma_\alpha P_R b ) (\bar \mu \gamma^\alpha P_R \mu)$) and  
$\Delta C^{\mu}_{9'} = -\Delta C^{\mu}_{10'} $ (which corresponds to the operator $(\bar s \gamma_\alpha P_R b ) 
(\bar \mu \gamma^\alpha P_L \mu)$) are clearly strongly disfavoured.

The other chiral operator $(\bar s \gamma_\alpha P_L b ) (\bar \mu \gamma^\alpha P_L \mu)$ 
(our hypothesis $\Delta C^{\mu}_9=- \Delta C^{\mu}_{10}$) turns out to be the closest to explain all the anomalies. 
However, even this operator fails to satisfy all the experimental results within their $1 \sigma$ ranges, in particular the value of 
$R_{K^*}^{\rm low}$.

The situation is slightly better for the operators involving electrons. 
It can be seen from Fig.~\ref{comb-e} that, while the primed operators are strongly disfavoured, the other two cases: 
$\Delta C^{e}_{9} = - \Delta C^{e}_{10} \approx 0.8$ and $\Delta C^{e}_{9} = \Delta C^{e}_{10} \approx -2$ 
work much better. In these two cases, $R_K$ and $R_{K^*}^{\rm central}$ can be satisfied within $1 \sigma$, and $R_{K^*}^{\rm low}$ within 
$\sim 1.5 \sigma$ and $\sim 1.3 \sigma$ respectively.

\begin{figure}[h!]
\begin{center}
\begin{tabular}{cc}
\includegraphics[scale=0.8]{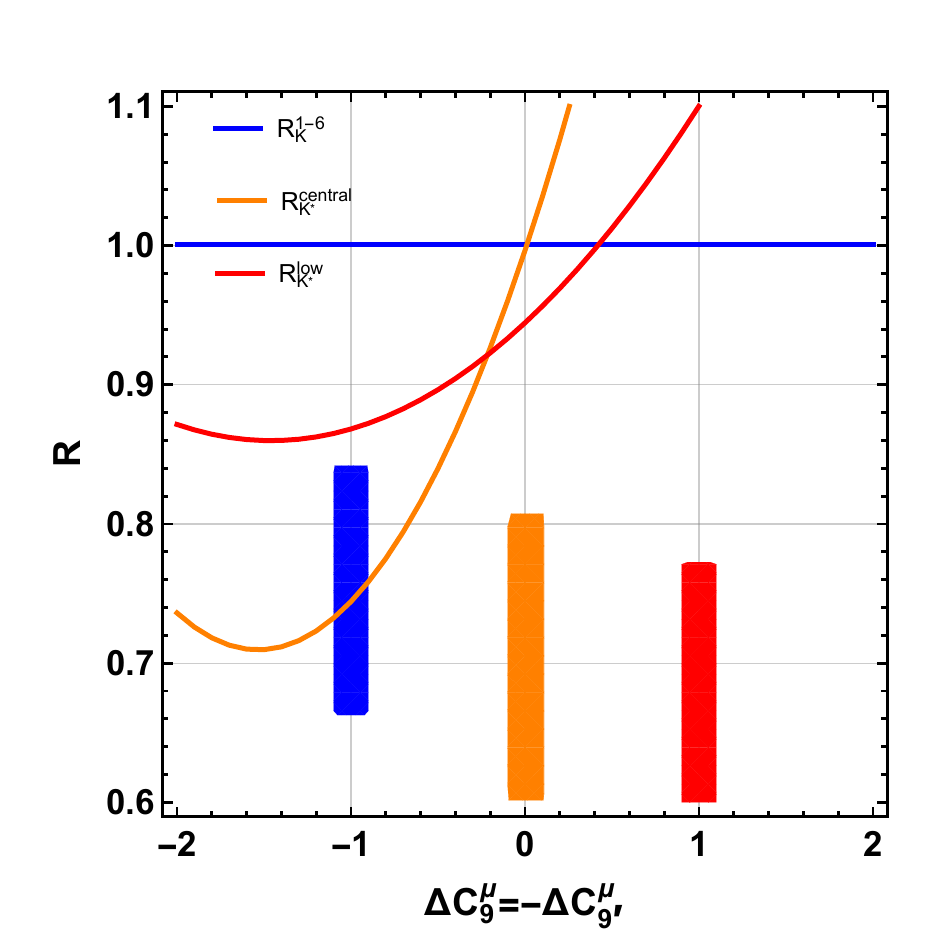} & \includegraphics[scale=0.8]{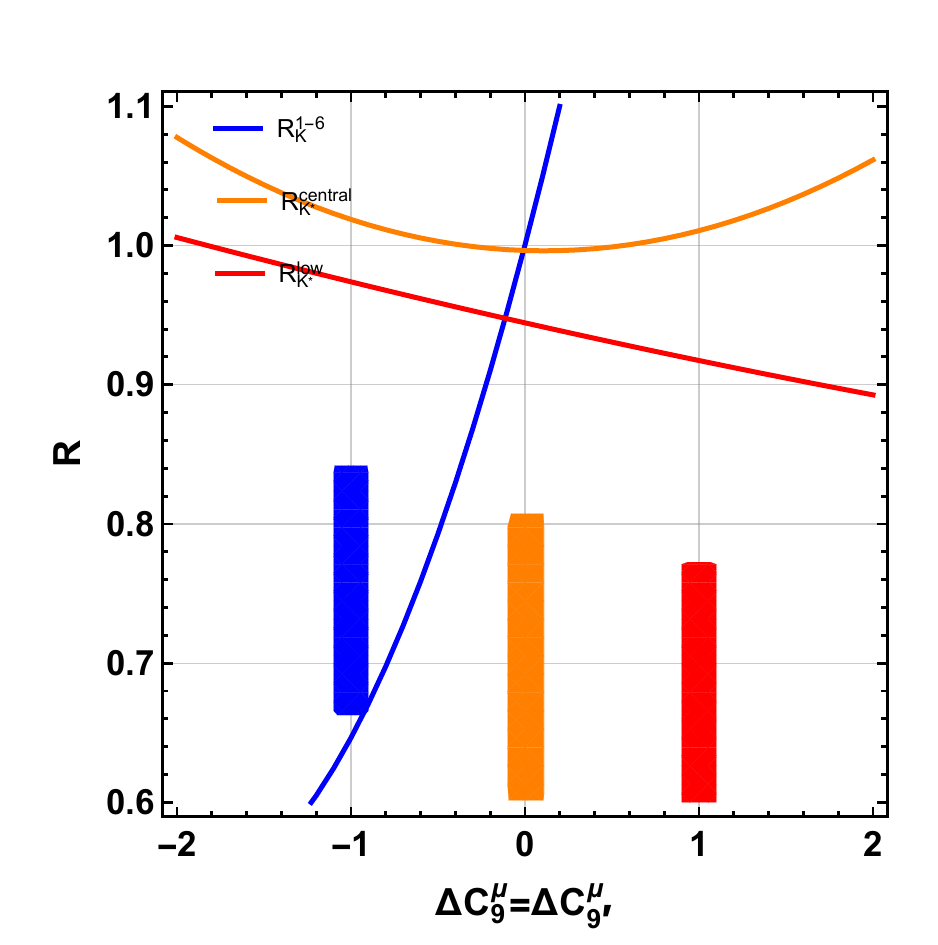} \\
\includegraphics[scale=0.8]{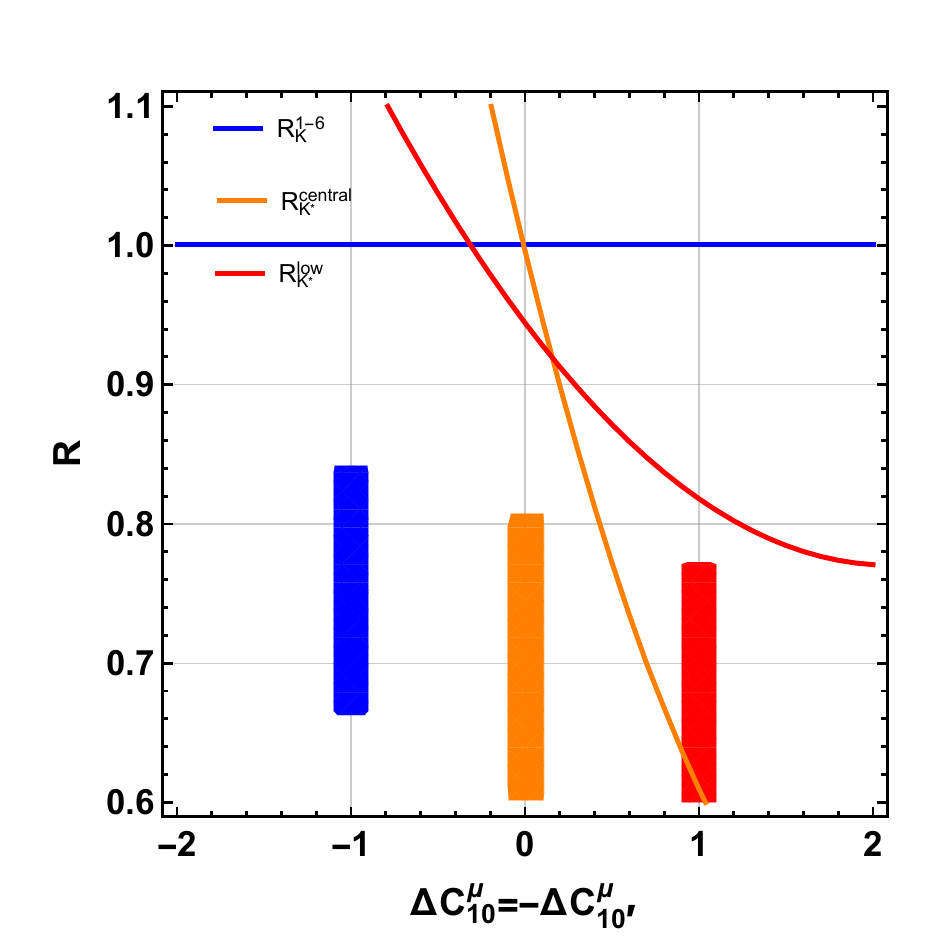} & \includegraphics[scale=0.8]{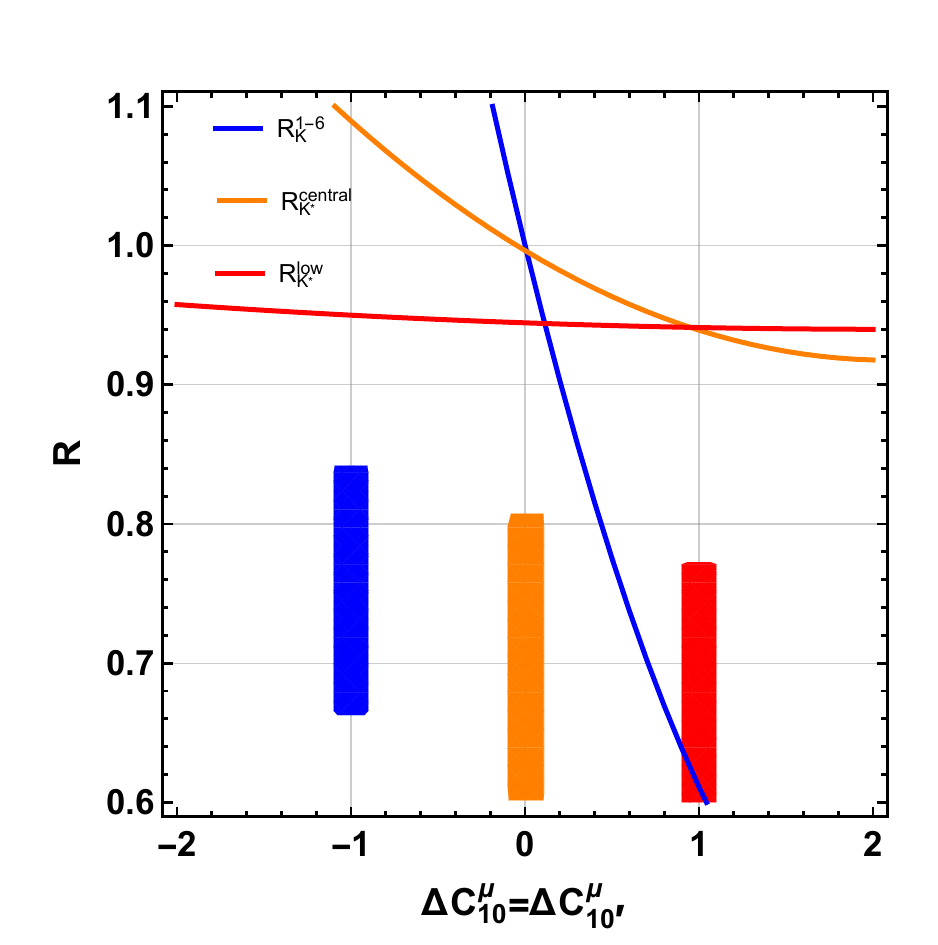}
\end{tabular}
\caption{Variations of $R_{K}$, $R_{K^*}^{\rm low}$ and $R_{K^*}^{\rm central}$ with the various vector and axial vector 
Wilson coefficients in the muon mode. The vertical bands correspond to the experimental $1 \sigma$ allowed regions 
(and independent of $\Delta C$). \label{comb2-mu}}
\end{center}
\end{figure}

\begin{figure}[h!]
\begin{center}
\begin{tabular}{cc}
\includegraphics[scale=0.8]{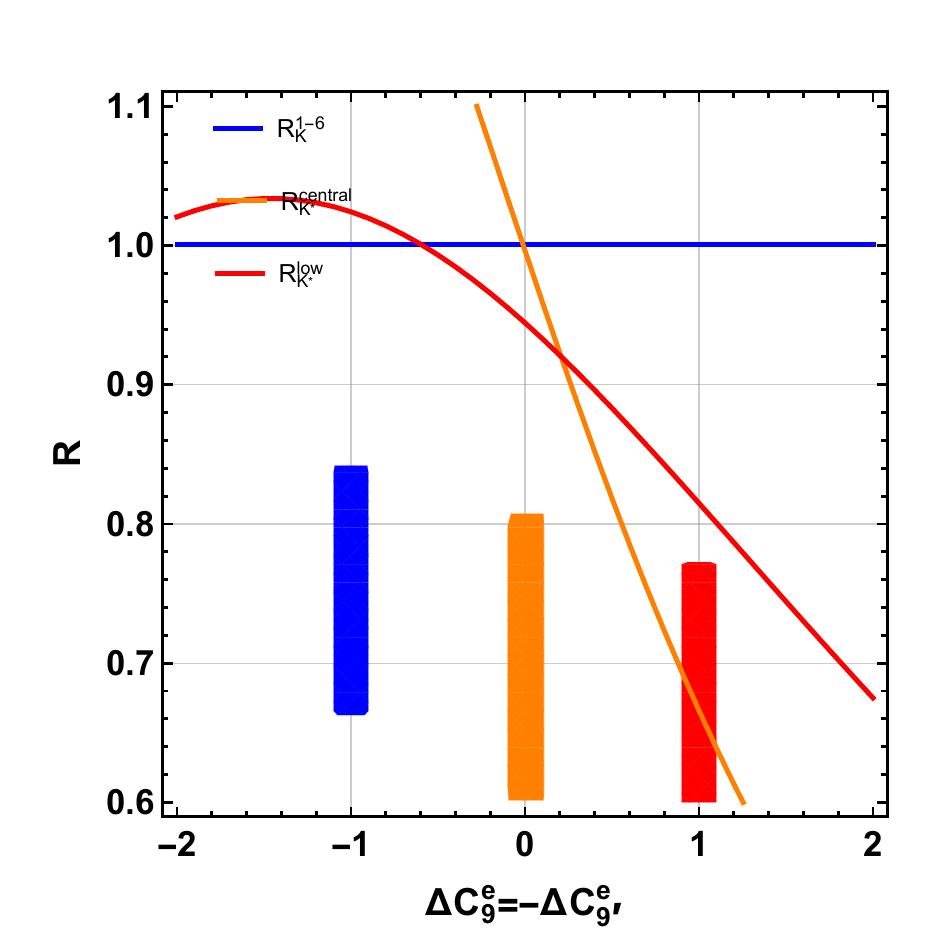} & \includegraphics[scale=0.8]{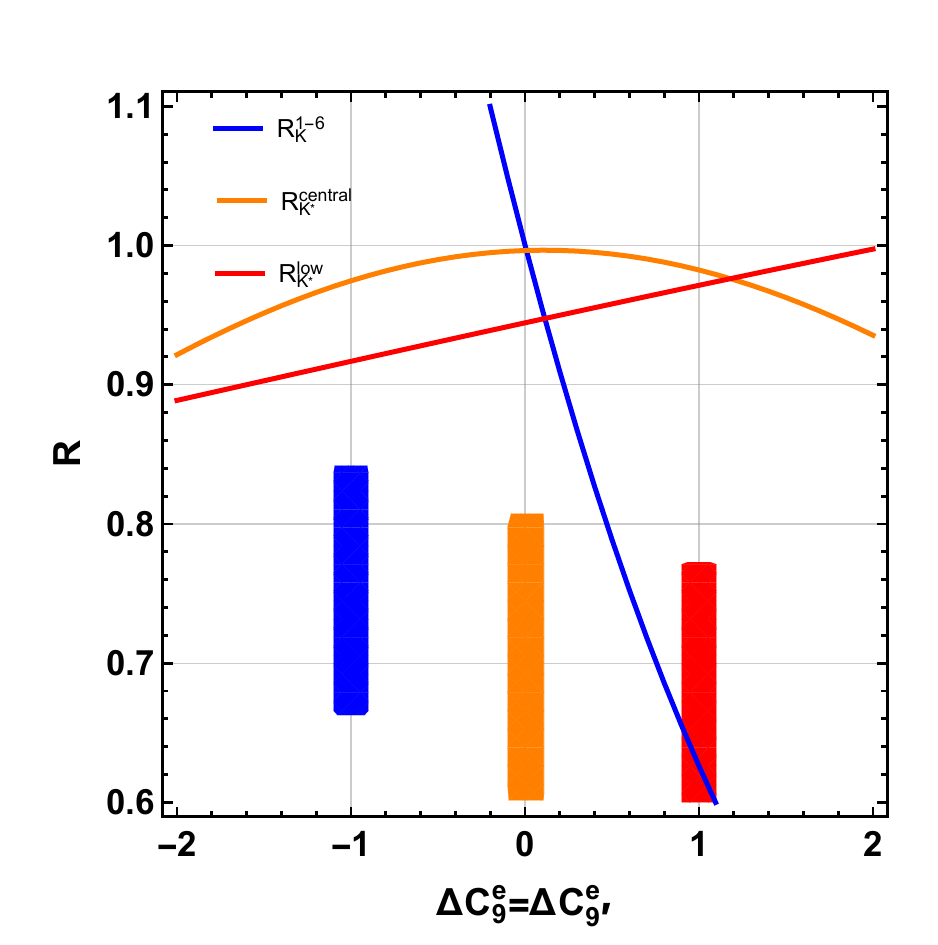} \\
\includegraphics[scale=0.8]{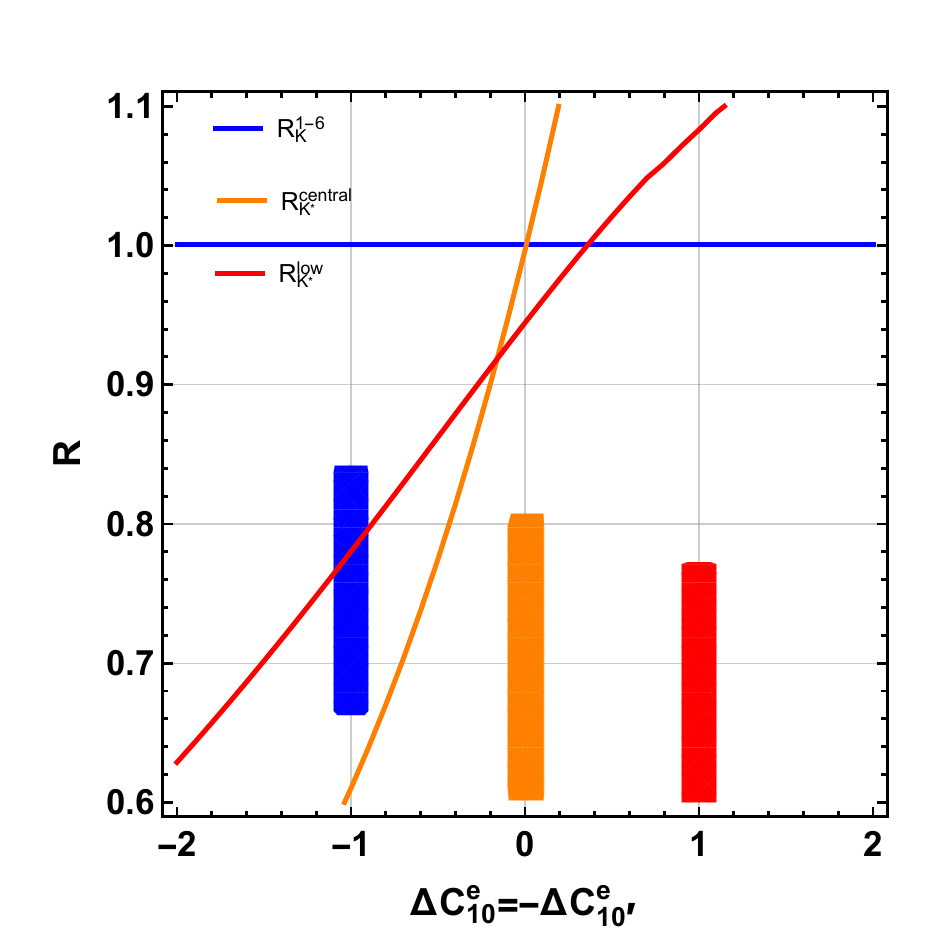} & \includegraphics[scale=0.8]{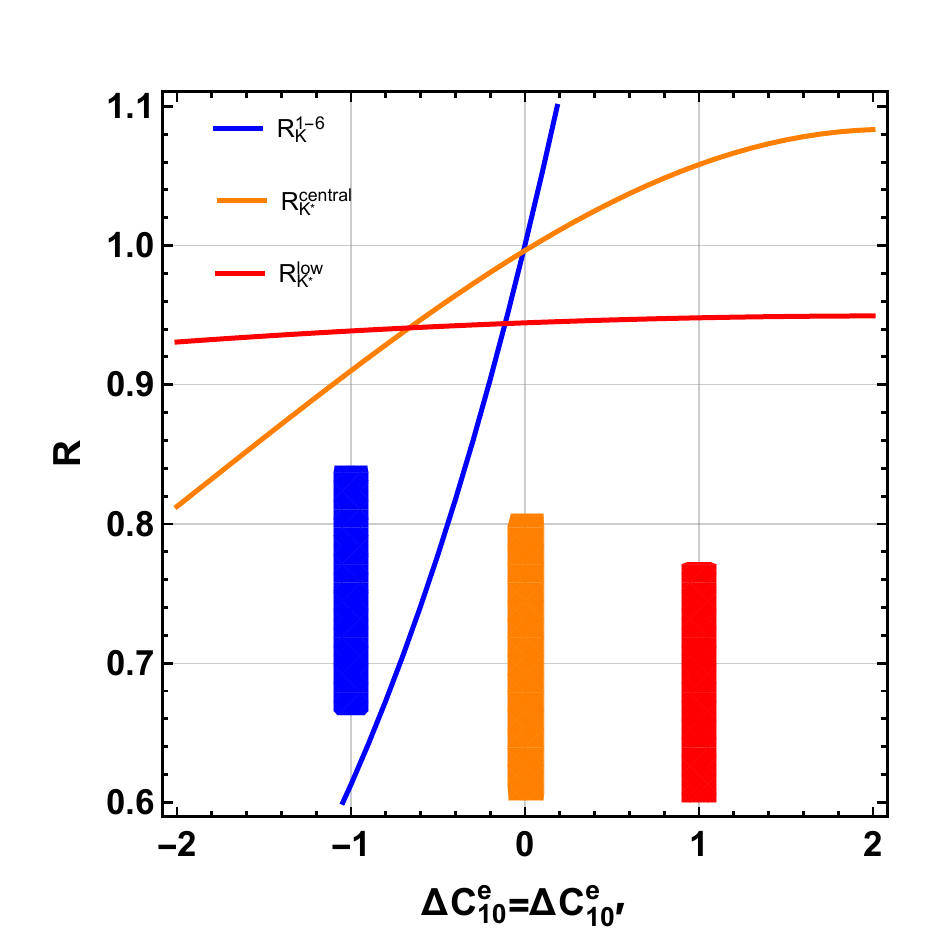}
\end{tabular}
\caption{Variations of $R_{K}$, $R_{K^*}^{\rm low}$ and $R_{K^*}^{\rm central}$ with the various vector and axial vector 
Wilson coefficients in the muon mode. The vertical bands correspond to the experimental $1 \sigma$ allowed regions 
(and independent of $\Delta C$). \label{comb2-e}}
\end{center}
\end{figure}

The scenarios  $\Delta C^{e,\mu}_{9} = \pm \Delta C^{e,\mu}_{9'}$ and $\Delta C^{e,\mu}_{10} = \pm \Delta C^{e,\mu}_{10'}$ are shown in 
Fig.~\ref{comb2-mu} and \ref{comb2-e}. It can be seen that they do not do a good job in explaining the anomalies simultaneously. 




Before closing this section, we would like to mention that we have also explored the possibility of existence of a pair of NP operators 
simultaneously with unrelated Wilson coefficients. For example, we have tried the following combinations: ($\Delta C_9^\mu, \Delta C_9^e$), 
($\Delta C_{10}^\mu, \Delta C_{10}^e$), ($\Delta C_9^\mu=-\Delta C_{10}^\mu, \Delta C_9^e=-\Delta C_{10}^e$), 
($\Delta C_9^\mu=\Delta C_{10}^\mu, \Delta C_9^e=\Delta C_{10}^e$) and all the 6 possible combinations 
($\Delta C_X^\mu, \Delta C_Y^\mu$) ($X,Y =9, 10, 9^\prime, 10^\prime$). 
However, even in these cases, we have not found solutions that explain $R_{K}$, $R_{K^*}^{\rm central}$ 
and $R_{K^*}^{\rm low}$ simultaneously within their respective $1\sigma$ allowed regions.

Hence, we conclude that, while local 4-Fermi operators of certain Lorentz structures (for example, $\Delta C^{e}_{9} = - \Delta C^{e}_{10} \approx 0.8$ as 
advertised above) can definitely reduce the tension with the SM considerably, they fail to bring all the 3 ratios within their experimental 
$1\sigma$ regions, in particular $R_{K^*}$ in the bin $q^2 =[0.045 - 1.1]$ GeV$^2$.

\subsection{Light vector boson to explain $R_{K^*}$ in the bin $q^2 =[0.045 - 1.1]$ GeV$^2$. }

Our investigations above show that local new physics  (i.e., $q^2$ independent Wilson coefficients) is unable to simultaneously 
explain  $R_{K}$, $R_{K^*}^{\rm central}$ and $R_{K^*}^{\rm low}$ at the $1 \sigma$ level. The main obstacle is to explain the result of 
$R_{K^*}$ in the low bin. This can be understood by noting that the branching ratio in the low $q^2$ region is dominated by the Wilson coefficient 
$C_7$ which is always lepton flavour universal. Quantitatively, in the $q^2$ bin $[0.045 - 1.1]$ GeV$^2$, the pure $C_7$ contribution constitutes 
approximately 73\% of the total branching ratio in the SM. On the other hand, the pure $C_7$ contribution is just about 16\% for 
the $q^2$ bin $[1.1 - 6]$ GeV$^2$. 

However, the situation can change in the presence of light degrees of freedom, for example, a very light ($\lesssim 20$ MeV) vector boson
$A^\prime_\mu$, with couplings 
\bea
\label{lag-light}
{\cal L} \supset - (\kappa_{bs} \, \bar b \gamma_\mu P_L s \, A^\prime_\mu + {\rm h.c.} ) - \kappa_{ee} \, \bar e \gamma_\mu P_L e \, A^\prime_\mu \, . 
\eea
 
\begin{figure}[h!]
\begin{center}
\begin{tabular}{c}
\includegraphics[scale=0.7]{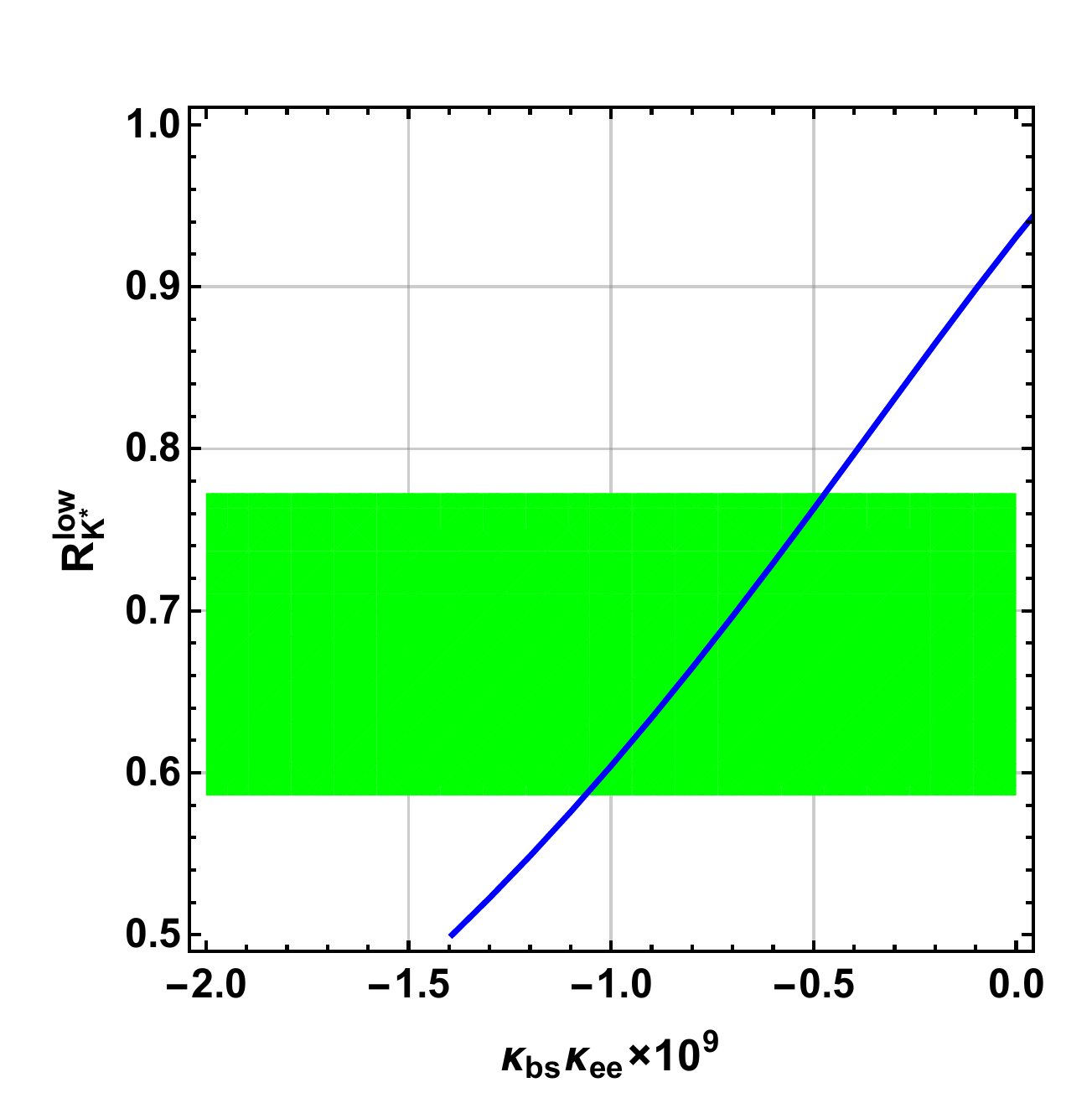} 
\end{tabular}
\caption{Variations of $R_{K^*}^{\rm low}$ with $\kappa_{bs} \kappa_{ee}$. The green band corresponds to the experimental $1 \sigma$ 
allowed region. We have used $m_{A'} = 17~\rm MeV$ in the numerical calculations. However, the result is not sensitive to the exact value of 
$m_{A'}$ as long as it is $\lesssim 50$ MeV. \label{kbskee}}
\end{center}
\end{figure}
 
The tree level exchange of the vector boson $A^\prime_\mu$ generates `$q^2$ dependent Wilson coefficients',  
\bea
\Delta C_9^e = - \Delta C_{10}^e &= & - \frac{1}{2} \bigg[\frac{4G_F}{\sqrt{2}}  \frac{\alpha_{\rm em}}{4 \pi}  |V_{tb} V_{ts}^{*}| \bigg]^{-1}  
\frac{\kappa_{bs} \kappa_{ee}}{(q^2 - m_{A'}^2) } \\
&=& - (6.15 \times 10^8) \frac{\kappa_{bs} \kappa_{ee}}{(q^2 - m_{A'}^2) \, [\rm in~GeV^2]}  \, .
\eea
The coupling combination $\kappa_{bs} \kappa_{ee} = -0.8 \times 10^{-9}$ generates $R_{K^*}^{\rm low}$ close to its experimental central value
\footnote{We find that a light gauge boson that couples to muons, instead of electrons, is unable to reproduce $R_{K^*}^{\rm low}$ below 0.8.}. 
In Fig.~\ref{kbskee}, we show how $R_{K^*}^{\rm low}$ varies with $\kappa_{bs} \kappa_{ee}$. It can be seen that 
$-1.1\times10^{-9} \lesssim \kappa_{bs} \kappa_{ee} \lesssim -0.5 \times10^{-9}$ is consistent with the experimental $1\sigma$ range of $R_{K^*}^{\rm low}$. 
We have also checked that values of $\kappa_{bs} \kappa_{ee}$ in the above range can be easily made consistent with the constraints 
coming from $\bar{B}_s - B_s$  mixing and anomalous magnetic moment of electron. 

However, the range $-1.1\times10^{-9} \lesssim \kappa_{bs} \kappa_{ee} \lesssim -0.5 \times10^{-9}$ generates $R_K$ and $R_{K^*}^{\rm central}$ 
in the range $0.89 \lesssim R_K, R_{K^*}^{\rm central} \lesssim 0.95$, well outside the experimental $1\sigma$ regions. 
Thus, separate local NP contributions, as discussed in the previous sections, are needed to explain $R_{K}$ and $R_{K^*}^{\rm central}$. 

We have checked that instead of a completely left-chiral coupling in Eq.~\ref{lag-light}, one can also use the following scenario
\bea
\label{lag-light-2}
{\cal L} \supset -(\bar b \gamma_\mu (\kappa_{bs}^L P_L + \kappa_{bs}^R P_R ) s \, A^\prime_\mu + {\rm h.c.} ) - \kappa_{ee}^V \, \bar e \gamma_\mu e \, A^\prime_\mu \, ,
\eea
which generates both $\Delta C_9^e$ and  $\Delta C_{9'}^e$, and works better than the previous case.  For example, 
($\kappa_{bs}^L \kappa_{ee}^V \approx -3.4 \times 10^{-9}$, $\kappa_{bs}^R \kappa_{ee}^V \approx -1.8 \times 10^{-9}$)  produces 
$R_{K^*}^{\rm low} \approx 0.67, R_{K^*}^{\rm central} \approx 0.93, R_{K} \approx 0.75$.

\section{Summary}
\label{conclusion}
In this paper, we have performed a model independent analysis of the recent LHCb measurements of $R_{K^*}$ in the two dilepton invariant mass bins 
$q^2 \equiv m_{\ell\ell}^2 = [0.045 - 1.1]$ GeV$^2$ and  $[1.1 - 6]$  GeV$^2$, along with an older measurement of a similar ratio $R_K$ in the pseudo scalar 
meson mode. 
We consider various possible $[\bar b \Gamma_\mu s] [\bar \ell \Gamma^\mu \ell]$ operator structures (both for the muon and electron modes), 
switching one operator at a time and also for specific combinations of them. We show that all the NP (pseudo) scalar operators and most of the (axial) 
vector operators are strongly disfavoured by the data. While some (axial) vector operators can explain $R_{K}$ and $R_{K^*}^{\rm central}$ at the same 
time, we found no operator that can explain all the three ratios (in particular, $R_{K^*}^{\rm low}$) simultaneously within their $1 \sigma$ 
experimental ranges. 

In order to explain also the $R_{K^*}^{\rm low}$, we proposed the existence of a very light ($\lesssim$ 20 MeV) vector boson with flavour specific 
couplings. We gave two examples shown in Eqs.~\ref{lag-light} and \ref{lag-light-2}. In the first case, we find that this new gauge boson, with couplings 
that explain $R_{K^*}^{\rm low}$, can neither explain $R_{K}$ or 
$R_{K^*}^{\rm central}$. 
Thus, additional local operators will be required to explain them together. As an example, a light gauge boson with coupling 
\bea
\kappa_{bs} \kappa_{ee} = -0.6 \times 10^{-9}
\eea
and additional local NP Wilson coefficients $\Delta C_9^\mu =  -\Delta C_{10}^\mu = -0.6 $ generates 
\bea
R_{K} = 0.69   \, , \, R_{K^*}^{\rm central} = 0.69\, \text{and} \, R_{K^*}^{\rm low} = 0.65 \, ,
\eea
all close to their experimental central values. 

In the second case, both  $R_{K^*}^{\rm low}$ and $R_K$ could be explained by the light vector only, however an explanation of  $R_{K^*}^{\rm central}$ 
as well would require additional, perhaps short distance, new physics. 

It remains a challenge to connect the existence of the light vector boson (with specific couplings) to heavy NP that generates the required 
short distance Wilson coefficients. We leave that for future work. 

We close with the comment that there might be issues with both the theoretical SM prediction (in particular, the uncertainty due to QED corrections) 
and the experimental measurement of $R_{K^*}$ in the low bin. In this work, we have taken the most recent SM prediction, the associated 
theoretical uncertainty and the experimental measurement at face value. Needless to mention that our conclusions may change if either of 
SM prediction/uncertainty  or the experimental measurement changes in future. 
\subsection*{Acknowledgement}
We thank Daniel Aloni for many useful discussions, comments on the manuscript and help with numerical calculations. 
Fruitful discussions with Ryosuke Sato, Masahiro Takimoto and Gilad Perez are also gratefully acknowledged. 
We also acknowledge the public code flavio \cite{david_straub_2017_555949} which has been used to cross-check 
some of the numerical results.

%
%
\providecommand{\href}[2]{#2}\begingroup\raggedright\endgroup

\end{document}